\documentclass[sigconf,9pt]{acmart}

\acmYear{2025}\copyrightyear{2025}
\acmConference[SenSys '25]{The 23rd ACM Conference on Embedded Networked Sensor Systems}{May 6--9, 2025}{Irvine, CA, USA}
\acmBooktitle{The 23rd ACM Conference on Embedded Networked Sensor Systems (SenSys '25), May 6--9, 2025, Irvine, CA, USA}
\acmDOI{10.1145/3715014.3722085}
\acmISBN{979-8-4007-1479-5/25/05}

\acmVolume{0}
\acmNumber{0}
\acmArticle{0}
\acmMonth{0}
\acmSubmissionID{843}

\usepackage{booktabs}
\usepackage{algorithm}
\usepackage{algpseudocode}
 \usepackage{multirow}
 \usepackage{tabularray}
\usepackage{amsfonts,balance,bbold,bm,color,hyperref,mathtools,microtype,multirow,soul,tabularx,xcolor,xspace}

\let\subcaption\relax
\usepackage{caption}
\usepackage{subcaption}
\usepackage{graphicx}
\usepackage{hyperref} 
\usepackage{graphicx}
\usepackage{verbatim}
\usepackage{amsfonts} 

\usepackage{tikz}

\usepackage{xcolor,colortbl}

\definecolor{Gray}{gray}{0.85}

\newcolumntype{a}{>{\columncolor{Gray}}c}

\newcommand{\isn}[1]{\sethlcolor{white}\hl{#1}}

\definecolor{azure}{rgb}{0.0, 0.5, 1.0}

\usepackage[title]{appendix}

\begin{document}

\title{Improving mmWave based Hand Hygiene Monitoring through Beam Steering and Combining Techniques}

\author{Isura Nirmal}
\affiliation{%
  \institution{University of New South Wales}
  \city{Sydney}
  \country{Australia}
}
\email{isura.bamunusinghe@unsw.edu.au}

\author{Wen Hu}
\affiliation{%
  \institution{University of New South Wales}
  \city{Sydney}
  \country{Australia}
}
\email{wen.hu@unsw.edu.au}

\author{Mahbub Hassan}
\affiliation{%
  \institution{University of New South Wales}
  \city{Sydney}
  \country{Australia}
}
\email{mahbub.hassan@unsw.edu.au}

\author{Elias Aboutanios}
\affiliation{%
  \institution{University of New South Wales}
  \city{Sydney}
  \country{Australia}
}
\email{elias@unsw.edu.au}

\author{Abdelwahed Khamis}
\affiliation{%
  \institution{Data61, CSIRO}
  \city{Brisbane}
  \country{Australia}
}
\email{abdelwahed.khamis@data61.csiro.au}

\begin{abstract}
We introduce \textbf{B}ea\textbf{M}steer\textbf{X} (\textbf{BMX}), a novel mmWave hand hygiene gesture recognition technique that improves accuracy in longer ranges (1.5m). \textbf{BMX} steers a mmWave beam towards multiple directions around the subject, generating multiple views of the gesture that are then intelligently combined using deep learning to enhance gesture classification. We evaluated \textbf{BMX} using off-the-shelf mmWave radars and collected a total of 7,200 hand hygiene gesture data from 10 subjects performing a 6-step hand-rubbing procedure, as recommended by the World Health Organization, using sanitizer, at 1.5m—over 5 times longer than in prior works. \textbf{BMX} outperforms state-of-the-art approaches by 31-43\% and achieves 91\% accuracy at boresight by combining only two beams, demonstrating superior gesture classification in low SNR scenarios. \textbf{BMX} maintained its effectiveness even when the subject was positioned $30^{\circ}$ away from the boresight, exhibiting a modest 5\% drop in accuracy.

\end{abstract}

\ccsdesc[500]{Human-centered computing~Gestural input}
\ccsdesc[500]{Human-centered computing~Ubiquitous and mobile computing systems and tools}
\ccsdesc[500]{Hardware~Beamforming}

\maketitle

\keywords{ACM proceedings, \LaTeX, text tagging}

\section{Introduction}
The availability of commodity mmWave radars in recent years has sparked significant research interest~\cite{10193776} into mmWave-based contact-less gesture recognition for a range of applications including hand hygiene monitoring~\cite{rfwash}, smartphone control~\cite{interactingwithsoli,jaime2016}, sign-language detection~\cite{santhalingam2020}, and so on. 
A key advantage of mmWave technology over cameras is its ability to operate effectively in challenging environmental conditions such as smoke, fog, and darkness, all while preserving user privacy—especially in healthcare settings. Compared to other radio technologies like WiFi~\cite{wifi2radar,9353723}, mmWave radars offer higher spatial resolution and more precise gesture recognition due to their wider bandwidth and co-located Tx-Rx antennas.

\isn{While the proposed system demonstrates comparable performance to RFWash}~\cite{rfwash} \isn{in hand sanitization applications, its primary advantages lie in enhanced sensing precision and privacy benefits though alternative technologies like acoustic (ultrasonic) sensors may provide a more cost-effective solution}.

\begin{table}[t]
  \caption{Review of mmWave Microgesture Recognition}
  \label{tab:mmwave-sensing}

  {\small
  \centering
  \begin{tabular}{|p{2.2cm}|p{1.2cm}|p{1.5cm}|p{2cm}|}
    \hline
    \textbf{Model} & \textbf{Sensing Range} & \multicolumn{2}{c|}{\textbf{Accuracy}} \\
    \cline{3-4}
    & & \textbf{Boresight} & \textbf{FoV-edge ($30^{\circ}$)} \\ 
    \hline

    CubeLearn\cite{zhao_lu_wang_trigoni_markham_2023} &$<$0.2m&90\%&Not tested\\
    RadarNet\cite{radarNet} &$<$0.2m&99\%&Not tested\\ 
    Soli\cite{jaime2016} &$<$0.3m&92.12\%&Not tested \\ 
    DeepSoli\cite{interactingwithsoli} &$<$0.3m&87\%&Not tested \\ 
    SolidsOnSoli\cite{solidsonsoli} &$<$0.3m&86\%&Not tested \\
    RFWash\cite{rfwash} &$<$0.3m&85\%&Not tested \\
    \hline
    \textbf{BMX} &$<$\textbf{1.5m}&\textbf{91}\%&>\textbf{86\%} \\
    \hline
  \end{tabular}

  }
\end{table}

However, current mmWave gesture recognition works have only explored scenarios with high signal-to-noise-ratios (SNR), which are typically exemplified by larger Radar Cross Section (RCS) gestures involving whole arm or leg movements~\cite{pantomime} (\textbf{Macro} gestures) in longer distances, or with very short radar-subject distances~\cite{interactingwithsoli} that minimize signal path loss. Thus, the gesture recognition capabilities of mmWave radars for detecting much smaller gestures (smaller RCS/\textbf{Micro} gestures) at longer distances, constituting \textbf{Low SNR Scenarios}, have yet to be explored (see Table~\ref{tab:mmwave-sensing}). 

\begin{figure}[htb]
    \centering
    \includegraphics[scale=0.73, trim=0 125 0 13, clip]{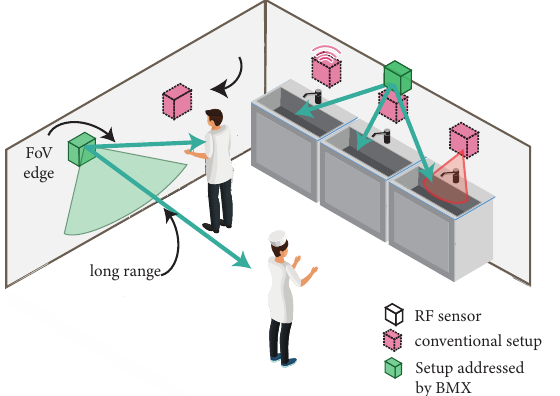}

    \caption{Environmental and application considerations can prohibit a \textit{user-facing} deployment setup (red) assumed by most RF sensing systems. Above  is an example of RF-based hand hygiene compliance monitoring. We propose \textbf{BMX} that uses special processing relying on Tx beamforming) to address a more practical yet challenging deployment setup (green). }
    \label{fig:bmx_motivation}

\end{figure}

\isn{Recognizing gestures from extended distances significantly enhances the scalability and versatility of mmWave radars. This advancement allows a single radar unit to serve a larger user base while expanding the range of feasible installation sites. It is especially advantageous in spacious environments, such as large rooms, where close proximity between users and devices is impractical, as illustrated in Figure}~\ref{fig:bmx_motivation}. \isn{Furthermore, minimizing dependence on a dense sensor network not only lowers infrastructure costs but also simplifies the integration of gesture recognition technology across diverse settings.}

This study examines the use of beamforming 
 
as a means to enhance gesture recognition in \textbf{Low SNR Scenarios}. Beamforming enables a radar with a linear antenna array to concentrate its radiation in a specific direction by adjusting the phases of individual antenna transmitting signals. By concentrating the transmitted power towards the gesture performer, beamforming can improve the SNR compared to non-beamforming scenarios, resulting in improved gesture detection performance.

Although beamforming is a well-known technique to increase SNR in a specific direction and thereby improve gesture sensing performance, applying it to mmWave radars can be challenging due to the antenna designs used in most commodity devices. These radars typically separate their transmit antennas by one wavelength ($\lambda$) to achieve time division multiplexing (TDM) multiple-input-multiple-output (MIMO) with independent Tx-Rx channels~\cite{ti_inc_2017}. However, when Tx beamforming is used with these devices, this large separation ($>\lambda/2$) and the limited number of Tx antennas (< 3) create side lobes that radiate significant energy towards directions other than the target of interest~\cite{petropulu2022}. As a result, the SNR for the main beam pointing towards the gesture may decrease, and if the energy of the side lobes is not fully utilized, the expected gains from beamforming may not be realized.

\textbf{BMX} \isn{introduces Beam Steering and Combining (BSC) to leverage side lobes for enhanced mmWave sensing. By steering multiple beams in different directions, even those not directly pointed at the gesture performer contribute to useful multipath structures}~\cite{ArrayTrack}. \isn{This allows the radar to capture gestures from multiple angles, enriching the sensing data. However, combining these beams is complex due to both constructive and destructive interference. To address this,} \textbf{BMX} \isn{employs deep learning to intelligently fuse echoed signals, outperforming traditional signal processing methods. As demonstrated in our evaluation in Section}~\ref{sec:evaluation},\isn{ BSC significantly enhances gesture detection accuracy compared to using a single beam.}

The contributions of this paper are summarized below:

\begin{itemize}
\item We propose a novel beamforming-based gesture detection system, \textbf{BMX}, suitable for micro gesture detection in \textbf{Low SNR Scenarios} using commodity mmWave radars. To the best of our knowledge,  \isn{micro-gesture sensing with mmWave radar—particularly with data-driven optimization} have not been previously investigated, hence \textbf{BMX} represents a humble step in this important research direction.
\item We introduce a self-attention based deep neural network to intelligently fuse the information from multiple beams to further enhance gesture detection accuracy beyond those achieved with a single beam pointed to the gesture performer. Our model features a novel data augmentation algorithm that takes the Doppler shift and multipath into account to improve the generalization of the model. 
\item We develop a \textbf{BMX} prototype using a Commercial-Off-The-Shelf (COTS) mmWave radar and evaluate it with a dataset of 12.4 billion analog-to-digital samples representing 7,200 hand rubbing (using a hand sanitizer) gestures performed by ten subjects following the World Health Organization (WHO) recommendations for maintaining hand hygiene. Our dataset was collected over a long-range, 5x longer than prior works. \textbf{BMX} outperforms state-of-the-art approaches by 30-43\% by combining only two beams. We will release the dataset to encourage further research in mmWave gesture sensing with beam steering. 
\end{itemize}

\section{Primer on mmWave Radar}
\label{sec:radar_fundamentals}
\begin{figure*} [htb]
 \centering
    \includegraphics[width = 1\linewidth]{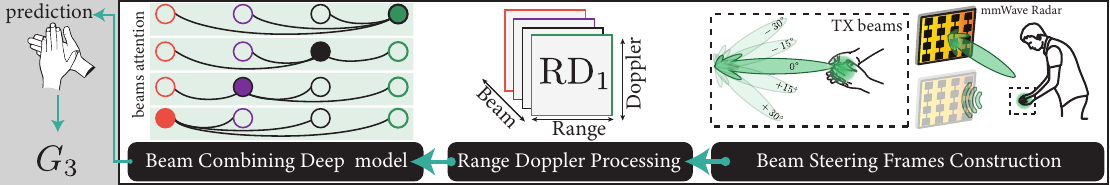}

    \caption{\textbf{\textbf{BMX} high-level system overview}. \textbf{BMX} combines advances in both radar signal construction  and deep learning to enable accurate recognition of intricate hand gestures (e.g., hand rub label $G_3$) under challenging deployment setups. } 
    \label{fig:mmwave_processing}   

\end{figure*}

 In this Section, we will briefly outline the signal processing steps to obtain range and velocity (Doppler) of an object monitored by FMCW radar, followed by its beamforming capability to focus radiation towards a specific direction. For more comprehensive details on these topics, readers can refer to \cite{richards2014fundamentals,xinrong2021}.

Range ($d$) can be determined by analyzing the time delay ($\tau$) of the received signal. Assuming a single reflector/single \textbf{Point Target/Object}, the time delay can be estimated from chirp-to-echo frequency difference $\Delta f_h$ according to: $\tau = \Delta f_h / S$, where $S$ is slope of the chirp. Consequently the reflector's range is $d = \frac{c\tau}{2}$, where $c$ is the speed of light and $\tau$ is the time delay between transmission and reception. In practice, however, multiple-point targets are expected, each with their respective chirp echo sent back to the receiver. To estimate frequency differences corresponding to multiple targets, a Fast Fourier Transform (FFT)  called \textbf{Range-FFT/1D-FFT} is conducted over a ``beat signal" (also known as \textbf{Intermediate Frequency}, or \textbf{IF} signal) formed by mixing transmitted and received signals\footnote{In fact, the mixed-signal will contain frequencies differences and sums as well. After low pass filtering, only the differences are retained.}.

The range resolution depends on the Bandwidth ($B$) of the radar as in~\cite{iovescu2017fundamentals}: 
\begin{equation}
    Range\; Resolution = c/ 2B
    \label{eq:range_res}
\end{equation}
With several GHz bandwidth, the range resolution in commodity mmWave radars is, therefore, in the order of a few centimeters only.

One limitation of Range-FFT output is that multiple objects, or multiple parts of a reflecting object, are indistinguishable when they yield the same range output. To overcome this limitation, and further improve the detection, mmWave radars can also detect the radial velocity, or the Doppler, of the target by transmitting multiple ($F$) chirps sequentially within a short time period, called a \textbf{Frame}. The sequence of phase values corresponding to a particular range bin from multiple chirps can be leveraged to estimate the velocity.

Eq.~\ref{doppler_equation} calculates Doppler Velocity $v$ from the phase difference ($\Delta \phi$) between two transmitted chirps where $T_c$ is the chirp duration.

\begin{equation}
\label{doppler_equation}
    v =\frac{\lambda \Delta \phi}{4 \pi T_c}
\end{equation}

For Doppler estimation within a complex scenario with many targets, the 1D range output of the $F$  
chirps within the whole Frame are stacked and a second FFT, called \textbf{Doppler-FFT/2D-FFT}, across the phase values of each column/range are taken. The outcome is a \textbf{2D Range Doppler Matrix (RDM)}, which can resolve multiple targets in the same range but with different Doppler velocities.

Modern radars use multiple transmitters and receivers to enhance spatial capabilities, particularly in angle estimation. An Rx antenna array with \( N \) elements and \(\lambda/2\) spacing achieves an angle resolution of \(2/N\). Adding more transceivers improves angular resolution but is costly due to the need for separate hardware pipelines.

\begin{figure}[tp]
\centering
\begin{subfigure}[t]{0.45\linewidth}
    \centering
    \includegraphics[width=\linewidth]{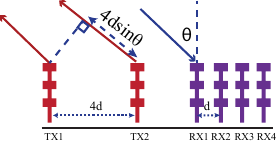}
    \caption{TDM-MIMO}
    \label{fig:mimo_arr}
\end{subfigure}\hfil
\begin{subfigure}[t]{0.45\linewidth}
    \centering
    \includegraphics[width=\linewidth,trim={0 0 0 0},clip]{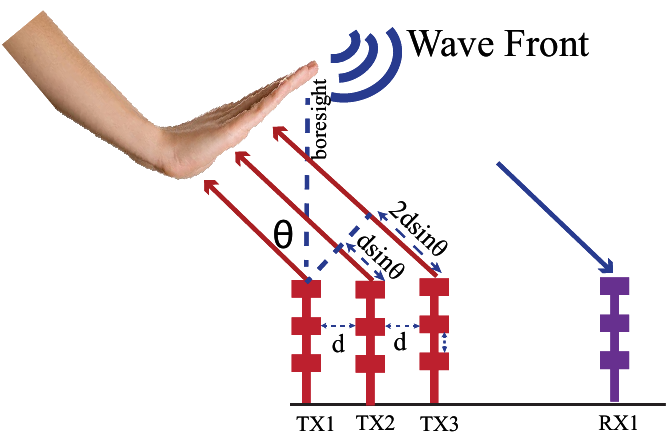}
    \caption[b]{Tx beamforming}
    \label{fig:beamforming}
 \end{subfigure}

 \caption{Concept of TDM-MIMO and Tx based beamforming}\label{fig:concepts}

\end{figure}

\isn{MIMO radar addresses this by using virtual antennas, enhancing spatial properties with existing hardware. The principle relies on linear phase changes between receiver antennas, allowing standard signal processing for angle resolution. As shown in Figure} \ref{fig:mimo_arr}, \isn{placing the second Tx at a distance of} \(4d\) \isn{from the first induces four times the phase change in the receiver array. Using TDM, where transmitters operate at different times, signals from different transmitters can be separated. This allows the signal-processing pipelines to utilize} \(N_{T_x \times R_x} = 2 \times 4\) \isn{IF data streams, effectively doubling the angular resolution with an additional Tx antenna.}

Beam steering, achievable through beamforming, dynamically changes the radar's beam direction. This enhances detection range in specific directions, utilizing phased arrays' phase shifting capability to rapidly switch between beams with microsecond latency.

\isn{\textbf{BMX} leverages Tx beamforming, which efficiently directs power toward the intended target, minimizing energy waste. Since Tx beamforming is implemented at the hardware level, it significantly reduces computational complexity in signal processing compared to Rx beamforming. This approach is particularly well-suited for long-range detection applications}~\cite{10663801}, \isn{making it an optimal choice for BMX.}.

Figure~\ref{fig:beamforming} depicts signal transmission involving three Tx antennas, Tx1, Tx2, and Tx3, spaced by $d$, transmitting towards a reflector at an angle $\theta$ from the boresight. Due to antenna spacing, signals from Tx2 and Tx3 travel additional distances of $d \sin \theta$ and $2d \sin \theta$ to the reflector compared to Tx1. To combine signals coherently, the radar applies phase shifts to compensate for signal arrival time differences. This ensures simultaneous signal reception and reflection, boosting SNR compared to no beamforming.

Varying phase vectors enable the radar to direct the beam to different angles. Within a frame, individual chirps can be configured with distinct phase vectors, referred to as a \textbf{Beam} in the upcoming sections, allowing rapid changes in beam direction. This capability is harnessed by \textbf{BMX} to achieve microsecond-latency beam steering in commodity mmWave radars. This facilitates obtaining multiple perspectives of a human gesture, as elaborated next.

\section{Motivation and Key Idea}

\label{sec:mulitipathOfBeamViews}

This section presents additional details to support and validate the central concept of directing the beam to multiple directions around the gesture performer, which could potentially enhance the accuracy of gesture classification.

\begin{figure}[t]
    \centering

    \includegraphics[width=0.7\linewidth]{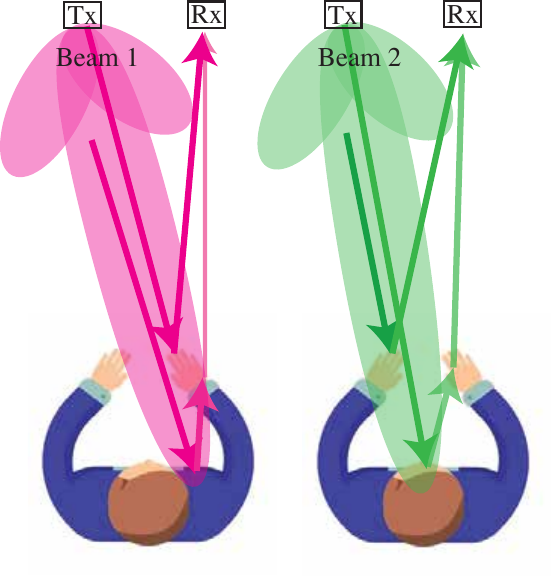}
    \caption{Different multi-paths created by different beams.}
    \label{fig:multipaths}

\end{figure}

Let us start by analyzing the composition of a received signal from a mmWave radar that steers a beam towards a gesture performer. When the beam's transmitted radio waves encounter the gesture performer, they will strike the performer at various locations and subsequently rebound back to the radar's receiver via distinct paths. The received signal, denoted by $y(t)$, can thus be expressed as the sum of these radio waves, each potentially experiencing different attenuation and phase: 

\begin{equation}
\label{eq:propagation}
    y(t) = 
    \sum_{m=1}^{M}A_me^{j\phi_m}e^{-j2\pi f\tau_m},
\end{equation}

where $M$ is the number of significant multi-paths between the transmitter and the receiver, $A_m$, $\phi_m$ and $\tau_m$ are the attenuation, phase and the delay of path $m$, respectively, and $f$ is the frequency.  Thus, the features extracted from $y(t)$  would have delay, phase and attenuation information of different radio propagation paths between a pair of radar transmitter and receiver, which can be explored for gesture sensing.

Let us now examine the scenario where two beams are directed towards the gesture performer (the location of the gesture performer is estimated through angle of arrival, or AoA, similar to~\cite{1449208}), but with a slight variation in their precise directions. This means that the phase shift vectors for the transmit antennas of the two beams differ slightly. Figure~\ref{fig:multipaths} illustrates this scenario, where the beams are steered sequentially within a very short time period (on the order of microseconds) to minimize the impact of gesture movement between the beams. As the radio waves from these beams hit different parts of the human body and hands, they produce different sets of multi-paths \isn{which aligns with the observations from}~\cite{ArrayTrack,10631013}. 

While a single beam suffices to encompass the human torso area (assuming a chest width of 40 cm < $\tan{15^0} \times 1.5m $), the signal intensity varies among these beams. Consequently, each distinct beam introduces unique multi-path effects. These various multi-paths yield distinct sensing data in terms of phase, attenuation and delay. Therefore, combining the signal features obtained from both beams can potentially enhance the overall sensing accuracy.

\begin{figure}[t]

    \begin{subfigure}[b]{0.22\textwidth}
        \includegraphics[width=\textwidth]{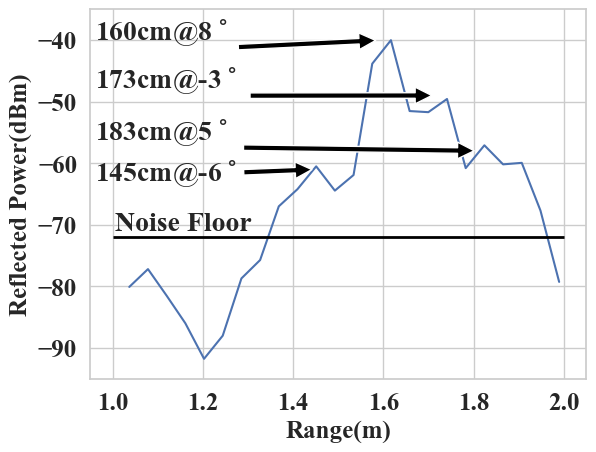}
        \subcaption{ Steered Beam @0$^\circ$}\label{fig:multipaths_2}
    \end{subfigure}
     \begin{subfigure}[b]{0.22\textwidth}
        \includegraphics[width=\textwidth]{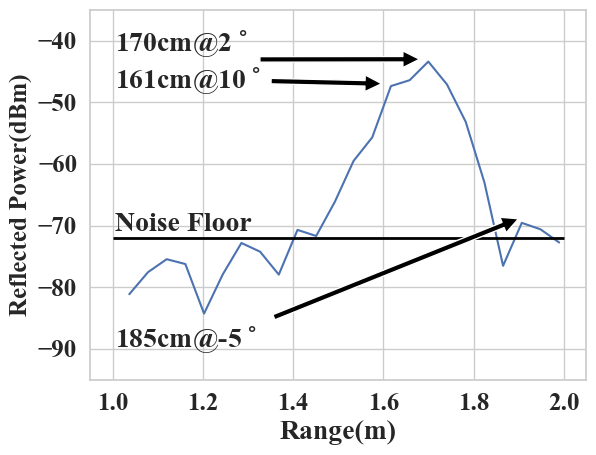}
        \subcaption{Steered Beam @15$^\circ$}\label{fig:multipaths_1}
    \end{subfigure}

    \caption{Range profiles for the same gesture observed by two different beams yielding different multipath structures.}\label{fig:range_profiles}

\end{figure}

To verify our hypothesis that different beams would produce distinct multi-path structures, we conducted a basic experiment in which we collected mmWave radar range profiles while a subject performed a particular gesture (the first gesture in the sequence shown in Figure~\ref{fig:gesture_set}). Figure~\ref{fig:range_profiles} displays the range profiles obtained from two different beams: one steered straight at the performer (0$^\circ$) and the other steered 15$^\circ$ to the left. The peaks in the range profile, along with their AoA (x cm @ y degrees) above the noise floor, correspond to significant multi-path reflections between the transmitter and receiver. These plots clearly demonstrate that even slightly rotated beams exhibit distinct multi-path structures, differing in number, range, and AoA, highlighting the impact of beam steering on signal propagation. \footnote{The range profiles are generated by the 1D-FFT signal processing introduced in Section~\ref{sec:radar_fundamentals}, while the AoA and
the noise floor by the method in~\cite{1449208} and the CFAR-CA method~\cite{richards2014fundamentals}, respectively.}

 \begin{figure}
    \centering
    \includegraphics[width=0.7\linewidth]{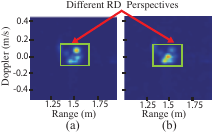}

    \caption{Range-Doppler images from different beam angles for the same gesture.}

    \label{fig:doppler_beams}
\end{figure}

To further validate that the information from multi-paths is available in the subsequent features derived from the received signal, we analyzed the RDMs of the same gesture captured from the two different beams, 0$^\circ$ and 15$^\circ$, as shown in Figure~\ref{fig:doppler_beams}. We observed distinct differences in the RDMs, indicating that the two beams provided us with two different perspectives of the same gesture. Combining these different perspectives may potentially complement each other, motivating us to investigate the use of beam steering and combining techniques for improved gesture recognition.   

\isn{Building on this motivation, the next section details how BMX applies beamsteering and signal combining to enhance gesture recognition accuracy.}

\section{BMX  Methodology}
\label{sec:BMX}

\isn{Commercial off-the-shelf (COTS) mmWave hardware is primarily designed with time-division multiplexed multiple-input multiple-output (TDM-MIMO) antennas, which are optimized for higher angular resolution}~\cite{ti_inc_2017}. \isn{While this design enhances resolution, it inherently restricts the beam steering coverage, typically within $\pm30^\circ$ (as confirmed by our experiments). Consequently, mmWave-based sensing through transmit (Tx) beam steering has been largely underexplored in the literature.}

\isn{To bridge this gap, we introduce BMX, a novel mmWave sensing framework that goes beyond TDM-MIMO by leveraging Tx beam steering in a structured and efficient manner. Unlike conventional approaches that focus on high-resolution virtual antenna arrays, BMX intelligently combines multiple beam measurements to enhance sensing accuracy. Our methodology integrates theory, simulations, and real-world experimentation to explore the full potential of Tx beam steering in mmWave sensing.}

\subsection{Transmit beamsteering with COTS mmWave radars}
\label{subsec:beamSteeringRadars}

\begin{figure}[t]
    \centering
    \includegraphics[scale=1]{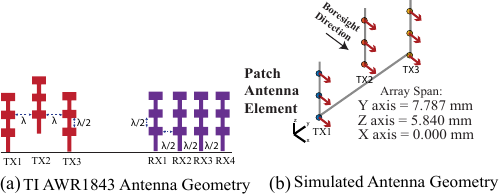}

    \caption{(a) AWR1843 antenna pattern,  and (b) 
    our simulation antenna array set-up.}

    \label{fig:geometry}
    
\end{figure}

Figure \ref{fig:geometry}(a) shows
the pre-printed patch antenna array geometry of a popular COTS radar, i.e.,
Texas Instrument's AWR1843 Boost~\cite{awr1843}, which 
is capable of performing Tx beam steering. Here, we first use computer simulations to understand the beam pattern for different TDM-MIMO and Tx steering modes of the AWR1843. 
The distance between the two Tx antenna elements is $\lambda$, where $\lambda \approx 4mm$. The 3 dB beamwidth of the azimuth plane is approximately ±28 degrees, and the elevation plane is ±14 degrees. 
Figure~\ref{fig:geometry}(b) illustrates the same patch antenna geometry we have created in our MATLAB-based simulation~\cite{MATLAB}.

\begin{figure*}[htp]
 \centering
    \begin{subfigure}[b]{0.16\textwidth}
        \includegraphics[trim={3.5cm 3.5cm 3.5cm 3.5cm},clip,width=\textwidth]{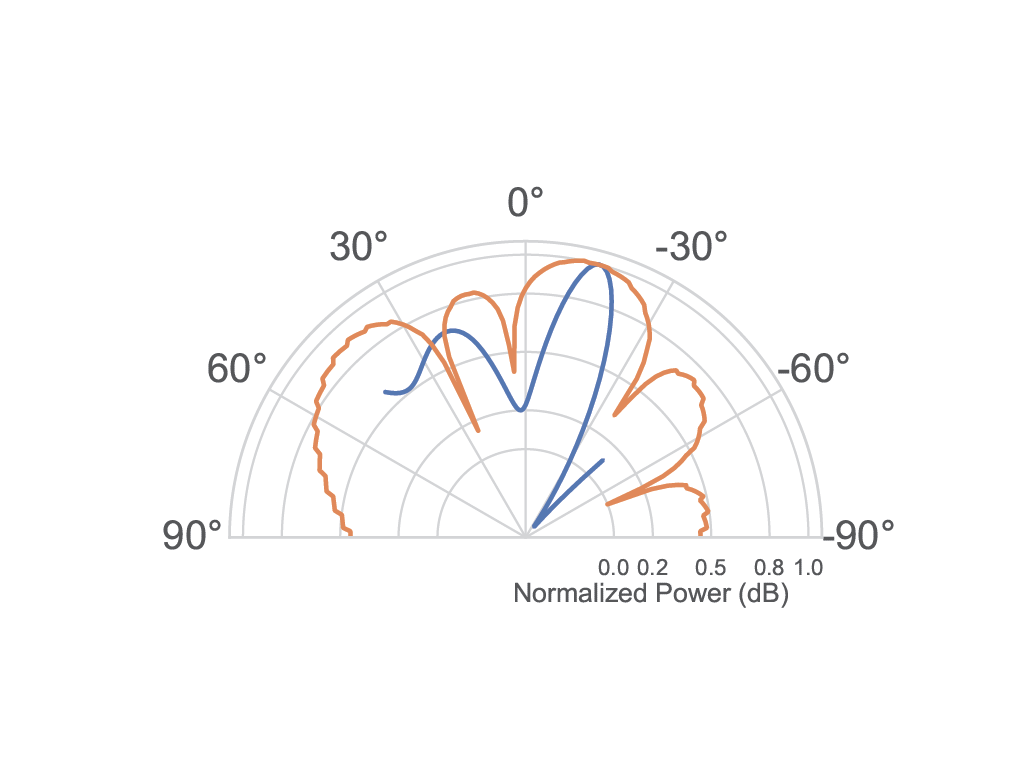}
        \subcaption{Beam @-15$^\circ$}\label{fig:sim_multipaths_1}
    \end{subfigure}
    \begin{subfigure}[b]{0.16\textwidth}
        \includegraphics[trim={3.5cm 3.5cm 3.5cm 3.5cm},clip,width=\textwidth]{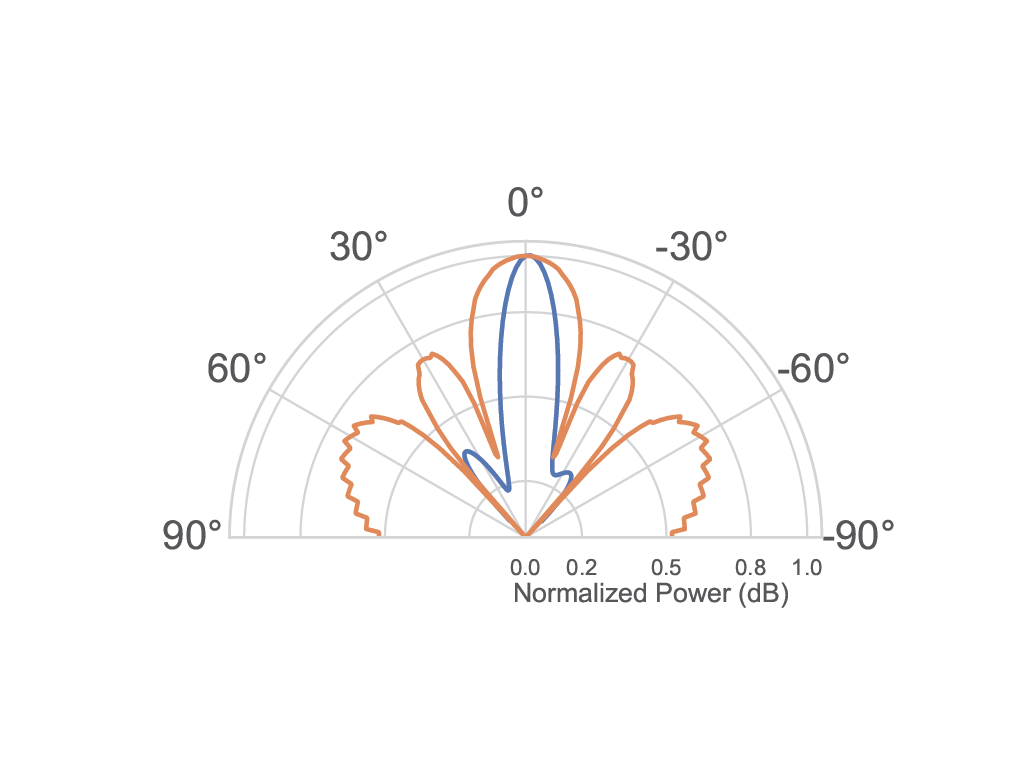}
        \subcaption{Beam @0$^\circ$}\label{fig:sim_multipaths_2}
    \end{subfigure}
        \begin{subfigure}[b]{0.16\textwidth}
        \includegraphics[trim={3.5cm 3.5cm 3.5cm 3.5cm},clip,width=\textwidth]{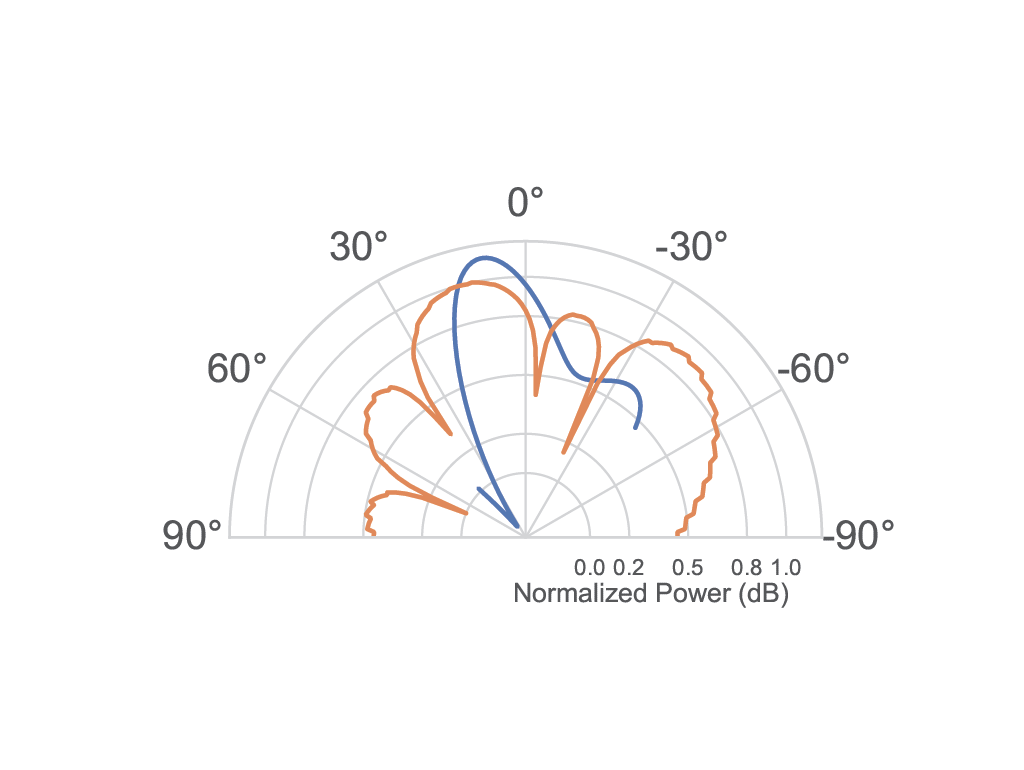}
        \subcaption{Beam @15$^\circ$}\label{fig:sim_multipaths_3}
    \end{subfigure}
    \begin{subfigure}[b]{0.16\textwidth}
        \includegraphics[trim={3.5cm 3.5cm 3.5cm 3.5cm},clip,width=\textwidth]{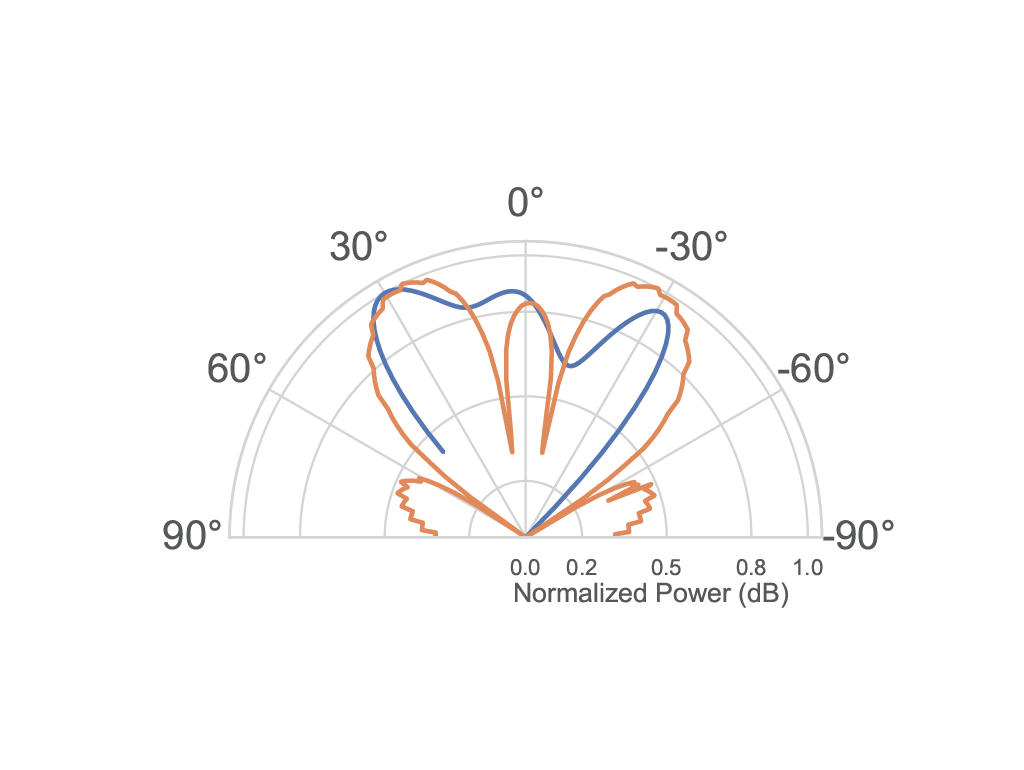}
        \subcaption{ Beam @$\pm$30$^\circ$}\label{fig:sim_multipaths_4}
    \end{subfigure}
        \begin{subfigure}[b]{0.16\textwidth}
        \includegraphics[trim={3.5cm 3.5cm 3.5cm 3.5cm},clip,width=\textwidth]{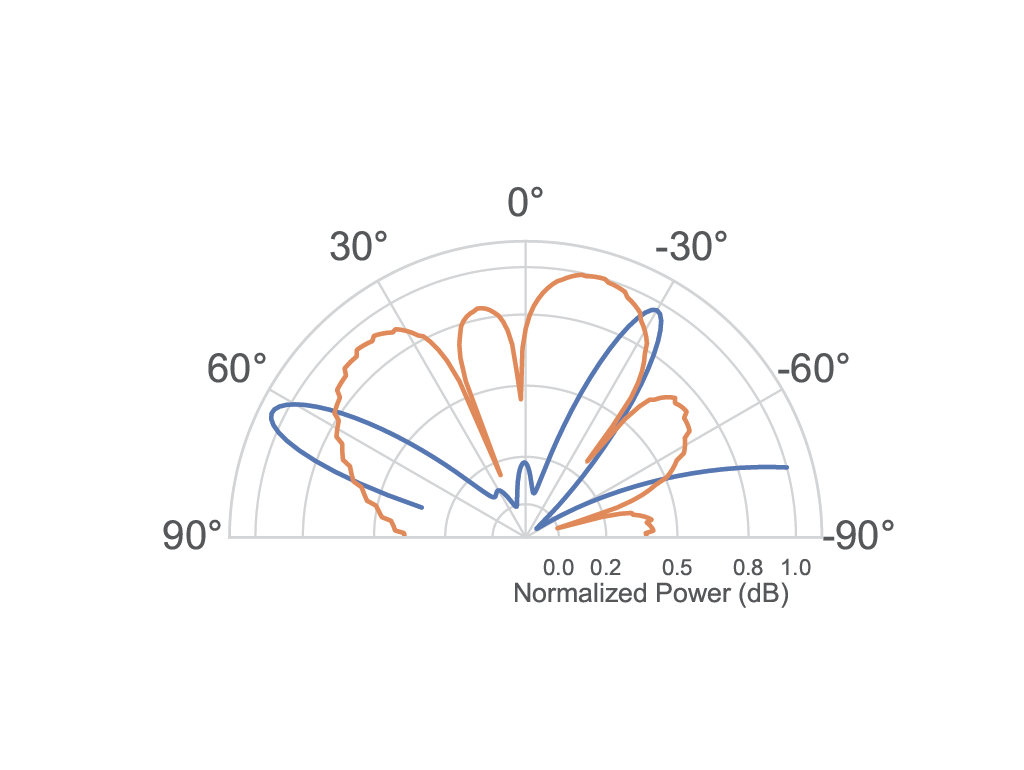}
        \subcaption{ Beam @45$^\circ$}\label{fig:sim_multipaths_5}
    \end{subfigure}
    \begin{subfigure}[b]{0.16\textwidth}
        \includegraphics[trim={3.5cm 3.5cm 3.5cm 3.5cm},clip,width=\textwidth]{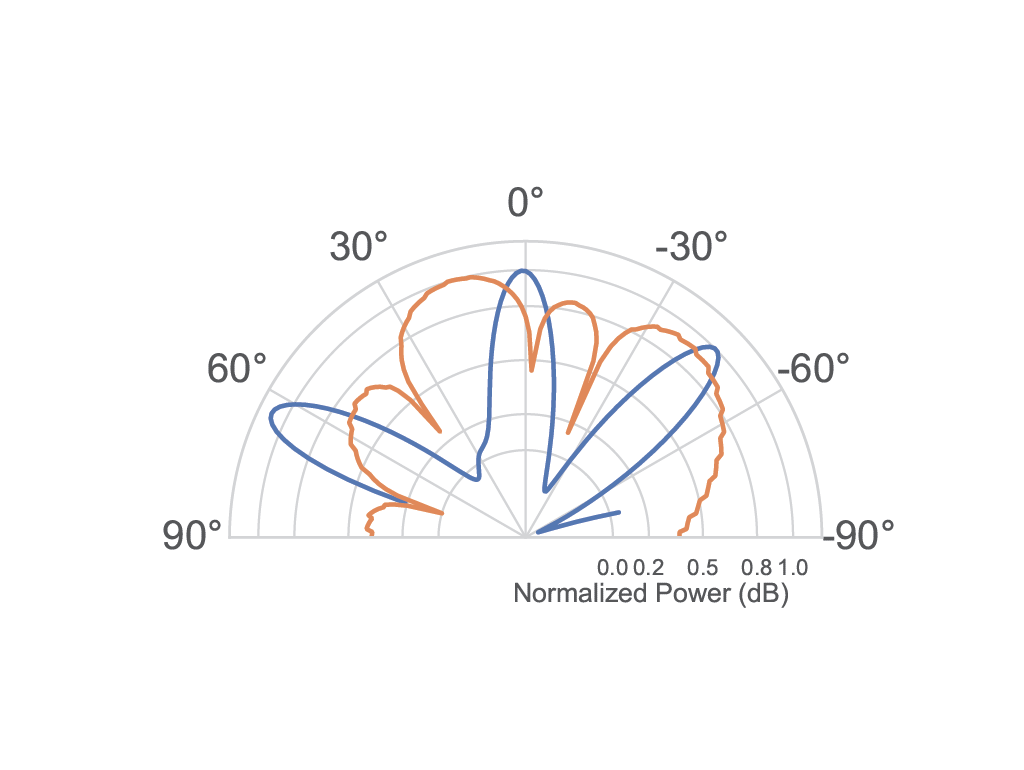}
        \subcaption{ Beam @-45$^\circ$}\label{fig:sim_multipaths_56}
    \end{subfigure}

    \caption{Simulated beam patterns (orange curves) and measured reflected power (blue curves) from  AWR1843}

    \label{fig:sim_matlab_bfpattern}
\end{figure*}

For the Tx beam steering modes, 
we first calculate the time-of-flight ($\tau{_N} = \frac{d_N \sin{\theta}}{c}$) for the additional distance ($d_N$) from the antenna element $N$ to a reflector (e.g., $d_1 = 0$, $d_2 = d$ and $d_3 = 2d$ in Figure~\ref{fig:beamforming}), where $\theta$ is the steering beam angle, i.e., the direction to focus the beam.

Then, the steering vector representing the relative phase shift values can be calculated as: 
\begin{equation} \label{eq:steering_vector}
\text{Steering Vector}_{N} = [e^{-j2\pi{f}\tau{_0}}, e^{-j2\pi{f}\tau{_1}},..., e^{-j2\pi{f}\tau{_{N-1}}}].
\end{equation}

For a radar that has $N$ antenna elements \textbf{optimized for better angular resolution when used as virtual antenna array} such as TI AWR1843, the distance between the first antenna and corresponding antenna ($d$) is usually configured as a factor of $\lambda$. 
 
Then, the phase shift vector $\phi_{N}$ for the antennas can be calculated by:
.
\begin{equation} \label{eq:phase_shifts}
\phi_{N} = [0, 2\pi \frac{d_2}{\lambda}\sin{\theta} , 2\pi \frac{d_3}{\lambda}\sin{\theta},...,2\pi \frac{d_N}{\lambda}\sin{\theta}].
\end{equation}

For example, AWR1843 has three Tx antennas (see Figure~\ref{fig:geometry}(a)), which has $\lambda$ spacing between two neighboring antennas, i.e., $d_1 = 0$, $d_2 = \lambda$ and $d_3 = 2 \times \lambda$. Therefore, the phase shift vector for AWR1843 may be simplified as:
\begin{equation} \label{eq:phase_shifts_AWR1843}
\phi_{N} = [0, 2\pi\sin{\theta}, 4\pi\sin{\theta}].
\end{equation}
Furthermore, it has a 64-step Tx phase shifter with a 5.625$^0$ step size, and  we can obtain the phase shifter values as:

\begin{equation} 
\label{eq:phase_shifter}
\phi^{'}{_N} = \phi{_N}/{5.625^\circ \times 64}
\end{equation}

Figure~\ref{fig:sim_matlab_bfpattern} plots the beam patterns TI AWR1843 obtained from the MATLAB simulations (orange curves) for different beamsteering angles ~\footnote{For $\theta = \pm30^\circ$, we obtain the same phase shift values from Eq.~\ref{eq:phase_shifts_AWR1843}. For the TDM-MIMO mode, where only one antenna is activated at a time slot (i.e., no beam-forming), our results

show a wider beam without any side lobes as expected, which is not
shown for brevity.} We observe that the beamsteering can increase the gain only within $\pm30^\circ$ because the side lobes become more powerful than the main lobe beyond 30$^\circ$. This happens due to the antenna design we mentioned earlier. Therefore, \textbf{BMX} will restrict beam steering within 30$^\circ$.

\begin{table}[t]
  \caption{AWR1843 Radar Configuration}
  \label{tab:radarconfig}
    {\small
    \centering
\begin{tabular}{|p{3.5cm}|p{4.2cm}|}
\hline
\textbf{Parameter}& \textbf{Value} \\
\hline
Device& TI AWR 1843 3 Tx: 4 Rx: 1 \\
Bandwidth& 3.61847 GHz (77 GHz - 80.68 GHz) \\
ADC samples per chirp & 256 \\
Number of Chirps & 40 \\
ADC sampling rate & 5,000 ksps \\
Frame rate & 16.6 frames/s \\
Tx power  & 12 dBM\\
Range Resolution& 4.14 cm\\
Velocity Resolution& 0.028 ms$^{-1}$\\
\hline
\end{tabular}
}

\end{table}

We further conduct an experiment in an anechoic chamber for the AWR1843 radar for multiple beamsteering angles ($\pm30^\circ$, 0$^\circ$, -15$^\circ$, +15$^\circ$,-45$^\circ$, +45$^\circ$) with the configuration in Table~\ref{tab:radarconfig}\footnote{For the Tx power, this is adjusted same for both TDM-MIMO and Beamsteering modes}. For each range bin, we plot the reflected power (blue curves in Figure ~\ref{fig:sim_matlab_bfpattern}) averaged over the 40 chirps sent within a frame. 

\isn{The simulated radiation pattern from MATLAB and the experimentally measured reflected power values at the receiver were normalized to a range of 0 to  to facilitate clearer visualization and comparison as follows.}

\isn{For the simulated radiation pattern from MATLAB, if} \( E_H(\theta, \phi) \) \isn{represents the horizontal component of the complex electromagnetic field in the radiation pattern, the normalized field pattern is given by} \( \left| \frac{E_H(\theta, \phi)}{E_{H,\text{max}}} \right| \).

 \( M \) \isn{represents the measured signal strength at} \( Rx_{1} \), \isn{and the normalized signal strength} \( M_{\text{norm}} \) \isn{is calculated as follows:} \( M_{\text{norm}} = \frac{M - \min(M)}{\max(M) - \min(M)} \).

\isn{Consistent patterns are observed between the simulated (orange curves) and measured power (blue curves). Despite conducting the experiments in an anechoic chamber, minor variations in the reflected signal were introduced due to hardware imperfections such as synchronization issues and mutual coupling.}

  \begin{figure*} [htb]
 \centering
    \includegraphics[scale = 0.8]{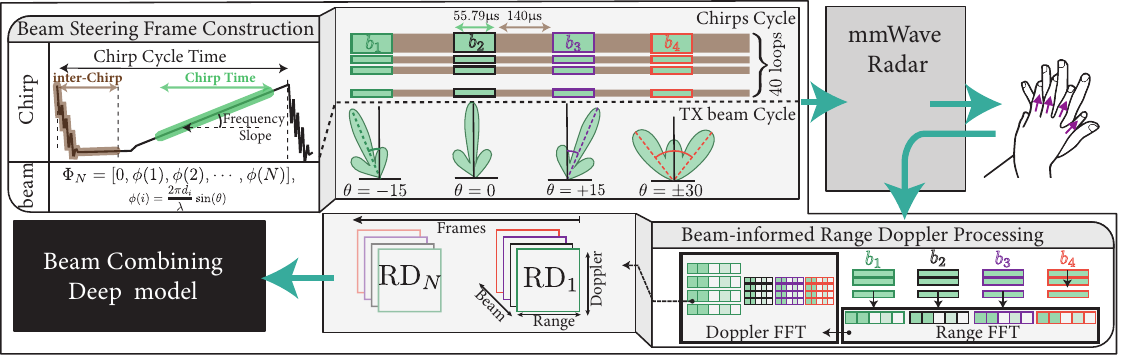}
 
    \caption{\textbf{BMX radar signal processing pipeline.}  } 

    \label{fig:radar_processing}    
\end{figure*}
 
. 
\isn{Our findings underscore a key distinction between BMX and conventional TDM-MIMO approaches:}

\begin{itemize}
    \item \isn{TDM-MIMO prioritizes fine angular resolution but lacks controlled Tx beam steering.}
    \item \isn{BMX exploits transmit beam steering to actively shape directional beams and enhance multipath sensing.}
    \item \isn{By fusing multiple Tx beams, BMX overcomes side lobe limitations and improves sensing accuracy beyond COTS radar capabilities.}

\end{itemize}

\isn{Through these innovations, BMX establishes a new paradigm in mmWave-based sensing, unlocking the untapped potential of Tx beam steering for real-world applications.}
\subsection{Signal Processing Pipeline}
\label{subsec:signalProcessingPipeline}
This section further explores the signal processing pipeline of \textbf{BMX}, which is shown in Figure~\ref{fig:radar_processing} (bottom).  Recall that the RDM is a data structure of \textbf{\#Range Bins} $\times$ \textbf{\#Doppler bins}. Each RDM is generated for each Tx beam, i.e., $b_1$, $b_2$,
$b_3$ and $b_4$ in Figure~\ref{fig:radar_processing} (bottom),  adding an extra dimension to the data structure. Depending on the number of frames within the gesture motion duration, we add another dimension called \textbf{Time Steps} to the data structure, 
i.e., Frames $RD_1$, $RD_2$,
$\cdots$ and $RD_N$ in Figure~\ref{fig:radar_processing} (bottom). Therefore, the dimensions of the final data structure, called \textbf{RDM sequences}, become: \textbf{Range Bins $\times$ Doppler Bins$\times$ Tx Beams $\times$ Time Steps}. We note that each element 
in the data structure is a complex representation of the reflected radio signal.

Utilizing our prior knowledge, assuming it is accessible through systems like millLoc~\cite{zhang2023}, regarding the subject's location and the correspondence of each range bin to a specific distance from the radar, \isn{we utilize a range-based filter to selectively isolate the range bins corresponding to the subject's potential location (e.g., filtering bins between 1.2 m and 1.8 m for a person positioned at 1.5 m). This approach not only reduces dimensionality along the range axis but also effectively suppresses background motion, enhancing the focus on relevant gestures within the sensing area.}

After generating the RDM sequences, we applied a background removal algorithm to remove the static reflections from the background. This is achieved by generating the mean of all the RDM along the time steps  and subtracting it from each RDM. We only retain the absolute values of the RDM maps since they correspond to the intensity of the reflected power. Finally, we further normalized the data values of the RDM data to be between 0 and 1, which is desirable for the deep learning method. 

Since the data collection involves multiple gestures performed by a diverse set of participants, the length of each gesture is different. However, the deep neural network requires us to use a fixed number of frames. Therefore, we calculated the minimum length of a gesture and applied undersampling to generate a unified-length data structure for our deep-learning model.

\begin{table}[t]
\centering
  \caption{Augmentation Configuration}
  \label{tab:augmentationconfig}
    {\small
    \centering
\begin{tabular}{|p{2.5cm}|p{4cm}|}
\hline
\textbf{Parameter}& \textbf{Range } \\
\hline
Distance &  from -0.369 m to 0.369 m \\
Doppler velocity&  from -0.14 m/s to 0.14 m/s  \\
\hline
\end{tabular}
}
\end{table}

\subsection{Data Augmentation}
\label{subsec:dataAugmentation}

\begin{algorithm}[htb]
{
\caption{Function for Augmenting Range Doppler Data}
\label{alg:augmentrange}
\begin{algorithmic}[1]
\Procedure{AugmentRangeDoppler}{$gesture$}
\State $a \gets$ a random doppler velocity value from Table~\ref
{tab:augmentationconfig}
\State $b \gets$ 
a random distance value from Table~\ref{tab:augmentationconfig}

\State $x \gets []$
\For{each \textit{gesture} $i$ in $gestures$}
\State $y \gets []$
\For{each \textit{beam} $k$ in gesture $i$}
\State $segment \gets$ a copy of beam $k$
\State velocities of $segment \gg a$ 
\State ranges of $segment \gg b$ 

\State $\text{noise mask} \sim \mathcal{N}(0, 0.01^2)$
\State  add the \text{noise mask} to $segment$ 

\State append the $segment$ to $y$
\EndFor
\State append the $y$ to $x$
\EndFor
\State \textbf{return} $x$
\EndProcedure
\end{algorithmic}
}
\end{algorithm}

\begin{figure}[htp]
    \centering
    \includegraphics[scale=1.1, trim=0 10 0 0, clip]{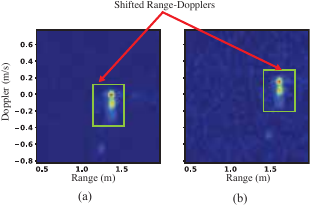}
    \caption{RDM example: before \& after augmentation. }
    \label{fig:augmentation_proc}
\end{figure}

Obtaining accurate and diverse RF data is a time-consuming and resource-intensive task while training data-driven deep learning models typically requires large datasets to achieve good performance. As a result, data augmentation has emerged as a popular method for increasing the number of samples and enhancing the generalizability of models.

To address this issue, we propose a novel data augmentation method for RF data, specifically RDMs, based on the radio channel effect including both multi-path and doppler shift.
Specifically, we augment the radio channel effect randomly by
changing the values of delay $\tau_m$ and phase $\phi_m$ in
Eq~\ref{doppler_equation} to represent the different
multiple path length and gesture radial/Doppler velocity ($v$). Table~\ref{tab:augmentationconfig} shows the
ranges of $\tau_m$ and $\phi_m$ values, which 
are chosen empirically. \isn{Ranges are chosen to be 9 times the range resolution (0.414 m) in both positive and negative directions from the original position. Similarly, velocities are selected as 4 times the velocity resolution (0.028 m/s) in both positive and negative directions. This radio data augmentation model is designed to handle variations in distance and velocity exhibited by the gesture performer, including differences in hand movement speed and slight body movements from the performer's location.}

\begin{figure*} [htp]
 \centering
    \includegraphics[scale = 0.8]{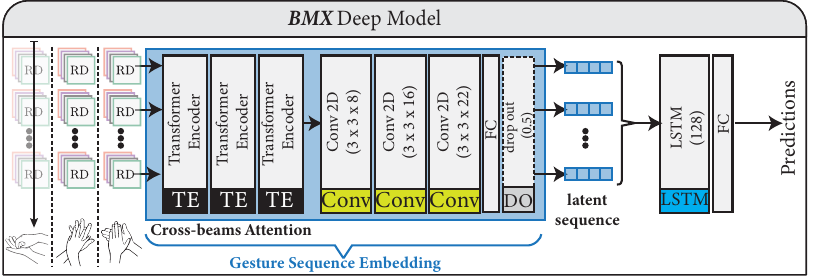}

    \caption{\textbf{BMX beam-combining and gesture classification deep model.}} 

    \label{fig:bmx_deep_model}    
\end{figure*}

Algorithm~\ref{alg:augmentrange} shows the augmentation algorithm where we randomly shift range and velocity bins, i.e., values (lines 9 and 10), from a predefined range in Table~\ref{tab:augmentationconfig} (lines 2 and 3) based on 
the real-world collected RDM. The amount of generated
augmented data is an empirically predefined number, which we
will evaluate in Section~\ref{subec:impactDataAugmentataion}.
We note that that $a$ and $b$ are real numbers, and negative values change the direction of \textbf{Shift Operator} $\gg$ in lines
9 and 10. Furthermore, we introduce an additional Gaussian noise mask to each augmented gesture at the end of the augmentation process to further diversify the augmented gestures (lines 11 and 12).

Figure~\ref{fig:augmentation_proc} shows the original (a) with its augmented counterpart (b) of an example RDM. 
Another way to interpret our data augmentation algorithm
is to move the gesture Doppler spectrum in the original RDM, e.g., the green box in
Figure~\ref{fig:augmentation_proc}(a), horizontally and 
vertically by changing the ranges, i.e., the lengths of multi-paths between the transmitter and receivers, and the
Doppler velocities, i.e., the speed of gesture motions, 
respectively, before adding random noise. Thus, incorporating a wider range of gesture velocities and distances into the training dataset can lead to significantly improved generalization of the trained models. This, in turn, can result in more robust gesture recognition outcomes.

\subsection{Deep Learning Model}
\label{subsec:dl_model}

Figure~\ref{fig:bmx_deep_model} shows the  multi-head attention-based deep learning model architecture
of \textbf{BMX} that fuses RDM from multiple beams to increase gesture recognition accuracy
since they have different multi-paths information caused by the gestures as discussed
in Section~\ref{sec:mulitipathOfBeamViews} earlier.

Here, the RDM dimensions are represented as follows: \textbf{Range Bins $\times$ Doppler Bins $\times$ Tx Beams $\times$ Time Steps } after the signal process pipeline discussed
in Section~\ref{subsec:signalProcessingPipeline} earlier. 
To begin the gesture classification process, for a batch of RDMs, we first iterate over the \textbf{Time Steps} dimension, applying multi-head attention layers to the input vectors and utilizing CNN layers to extract spatial information from each RDM frame. \isn{ 
The Tranformer Encoder based self-attention}~\cite{attention} \isn{layers dynamically assign higher attention weights to crucial features, such as constructive multipath components within each beam, while suppressing less relevant or destructive information. This mechanism enables the model to selectively enhance the most informative signal reflections, improving the overall effectiveness of beam combination and gesture recognition.}

After the \textbf{Gesture Sequence Embedding}, we concatenate the \textbf{latent sequence} and use them for temporal feature extraction using LSTM layer. Finally, the output is passed through a softmax layer for predicting the gesture represented by the input RDM.

Let $V$, $K$ and $Q$ be input vector, key vector and query vectors respectively (with dimensionality equal to \textbf{Range Bins $\times$ Doppler Bins}). Then,  the dimension of the key vector $K$ is $\sqrt{d_k}$ .   
.
A Transform Encoder attention layer can be modeled as:
\begin{equation}
\label{alg:attn_1}
Attention(Q, K, V) = softmax(\frac{QK^T}{\sqrt{d_k}})V,
\end{equation}
where $T$ is the transpose operation.
We note that $K$ and $Q$ are initialized internally based on the given $V$. Then, a single head operation $head_i$  can be represented as:
\begin{equation}
\label{alg:attn_2}
head_i = Attention(Q W^Q_i, K W^K_i, V W^V_i),
\end{equation}
where $W^Q \in \mathbb{R}^{d_{\text{model}} \times d_k}$, $W^K \in \mathbb{R}^{d_{\text{model}} \times d_k}$, $W^V \in \mathbb{R}^{d_{\text{model}} \times d_v}$ are parameter matrices. Finally, the operation of the multi-head attention layer
of $h$ head is:
\begin{equation}
\label{alg:attn_3}
MultiHead(Q, K, V) = Concat(head_1, ..., head_h)W^O,
\end{equation}
where $W^O \in \mathbb{R}^{hd_v \times d_{\text{model}}}$ is another parameter matrix.

The output of three Transformer Encoder attention layers
will then be input to three 2D Convolutional Neural Network (CNN)
layers
with different numbers of filters to extract
gesture spatial features as shown in Figure~\ref{fig:bmx_deep_model}.

The output of CNN layers will be further passed through a Fully Connected (FC) layer and a Dropout layer before being concatenated  into a latent sequence and passed to the  temporal feature extraction module, which consists of a 128 time-step Long Short-Term Memory networks (LSTM)~\cite{hochreiter1997long} and another
FC layer. Finally, the output of the FC layer is passed to
a Softmax to make the final hand gesture label prediction.

\section{Evaluation}
\label{sec:evaluation}

\subsection{Goals, Metrics and Methodology}
We will evaluate \textbf{BMX} with a critically  important
hand gesture recognition application, hand hygiene monitoring. 

\subsubsection{Hand hygiene monitoring}

\begin{figure}[htp]
    \centering

    \includegraphics[width = 1\linewidth]{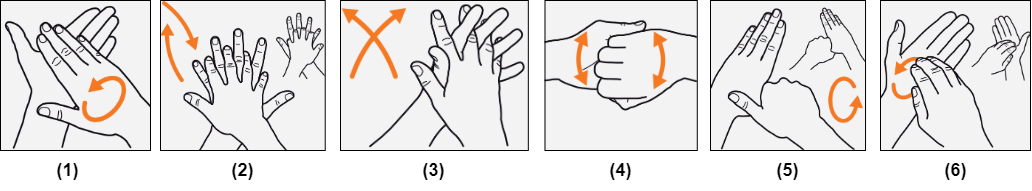}

    \caption{\textbf{Hand Rubbing Procedure as recommended by the WHO.} Hand Rub,\textbf{(1)}  Palm to Palm with only one hand moving \textbf{(2)} Right Palm over left dorsum with interlaced fingers and vice versa where only one hand moving at a time \textbf{(3)} Palm to Palm with interlaced fingers where both hands move \textbf{(4)} Back of fingers to opposing palms with fingers interlocked where both hands and forearm moving\textbf{(5)} Rotational rubbing, of left thumb, clasped in the right palm (stationary) and vice versa \textbf{(6)} Rotational Rubbing where left-hand stationery. (Image Courtesy\cite{whohandrub}) }
    \label{fig:gesture_set}
\end{figure}

Even before the COVID-19, Healthcare-Associated Infections (HAIs) seriously impact patient safety and healthcare costs. In the EU, there are about 2.6 million HAIs annually, causing 91,000 deaths, while the USA reports 1.7 million HAIs leading to 99,000 deaths.
{Regions with low-income countries face more challenges, with limited data highlighting higher HAI rates and significant health/economic consequences}~\cite{dekraker2022}. Improving hand hygiene compliance by following
the hand rubbing using sanitizer procedure in Figure~\ref{fig:gesture_set} is 
the simplest way to reducing HAIs according to the WHO.

\isn {Accurate tracking can enhance compliance verification, ensuring that healthcare workers adhere to hand hygiene protocols—critical for infection control and preventing the spread of pathogens in medical environments. One key advantage of long-range sensing is its ability to monitor hand hygiene compliance across larger areas with fewer sensors, reducing deployment costs while maintaining comprehensive coverage. This capability is particularly beneficial in high-traffic healthcare settings where continuous monitoring from various angles is necessary. Additionally, improved tracking provides valuable data for public health outcomes, enabling better analysis of infection trends and informing proactive intervention strategies.}

However, detecting such hand rubbing gestures accurately using mmWave radar is extremely challenging due to their small RCS. RFWash~\cite{rfwash} recently addressed this challenge, but its sensing coverage is limited to within 30 cm of the radar's bore sight angle, increasing system cost for covering a given Area of Interest (AoI). Unlike handwashing, which requires specific facilities, sanitizers can be conveniently installed or placed anywhere as illustrated in Figure~\ref{fig:bmx_motivation}. With sanitizers, individuals aren't tethered to a wash basin; they can apply the sanitizer and then move further away. This poses a challenge to traditional mmWave-based sensing systems, which struggle to maintain proximity and orientation. Additionally, the flow of water does not impact the radar signal reflection, as the process does not involve washing hands with water. \textbf{BMX} overcomes this limitation by intelligently fusing information from multiple transmit beams at different angles, expanding the sensing range of a radar to \isn{1.5m, which is five times more than that of RFWash,  and its coverage area by up to 25 times, resulting in an order-of-magnitude system cost saving}.

\subsubsection{Data Collection}

\begin{figure}[htp]
    \centering
    \includegraphics[width=0.6\linewidth, angle =90 ]{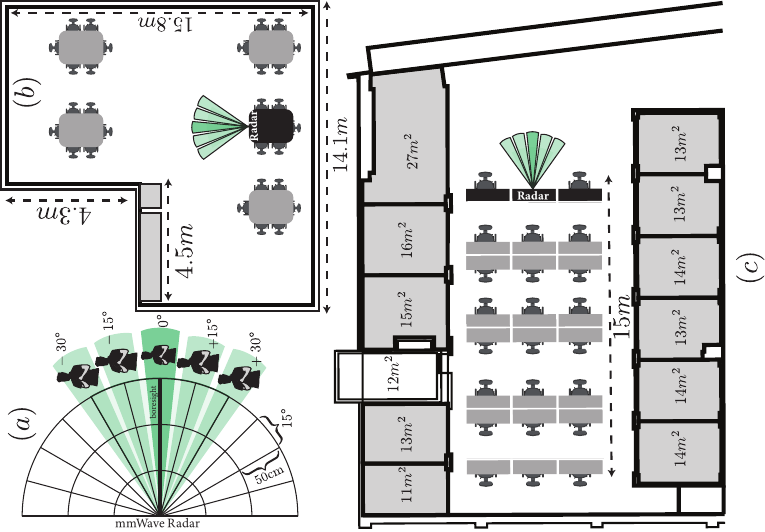}

    \caption{\textbf{BMX data collection setup}. (a) radar deployment setup and user positions.  (b) and (c) are the layouts of the data collection environments in a meeting room and an open plan office respectively. }
  
    \label{fig:data_collection_env}
    
\end{figure}
\isn{Ten participants (5 male, 5 female) with an average weight of }63.5$\pm$17.37 kg, height of 167.5$\pm$8.9 cm, and age of 27.4$\pm$3.5 years\footnote{UNSW Human Ethics HC17823 approved our data collection.} performed six gestures at five locations, spaced 15$^\circ$ apart from -30$^\circ$ to 30$^\circ$ relative to the radar, at a distance of 1.5m. \isn{The study was conducted in two distinct environments: an open-plan office with occasional human movement and a meeting room with minimal disturbances. Figure}~\ref{fig:data_collection_env} \isn{illustrates the data collection layouts. Each participant performed the gestures 12 times per location, ensuring a diverse and representative dataset. To account for variability, participants differed in body sizes, arm lengths, and hand motion speeds, with each gesture taking 3 to 6 seconds to complete.)}

\isn{Standardized instructions from an instructional video and the data collector ensured consistency across participants, improving the generalizability of the collected data to real-world settings.}

Each gesture was captured using 42 radar frames, each with 160 chirps (40 chirps per beam for 4 beam angles per frame at -15$^\circ$, 0$^\circ$, 15$^\circ$ and 30$^\circ$. Recall that $\pm30^\circ$ beam angles produce the same phase shift values as discussed in
Section~\ref{subsec:beamSteeringRadars} earlier). From each chirp, we collected 256 ADC samples (using the same configuration presented in Table~\ref{tab:radarconfig}). Thus, we collected 12.4 billion ADC samples in total, representing 7,200 gestures. We also collected TDM-MIMO data by inserting extra chirps within the same frame~\footnote{We configured the antenna transmit powers for each chirp so that the total transmit powers for both the TDM-MIMO and the beamsteering mode are equal to 12dBm.}. We will release the dataset to stimulate further research in this area once it is published.

\subsubsection{Deep Model Training Parameters}

AdamW\cite{Loshchilov2017DecoupledWD} optimizer was used for optimization with an initial learning rate of $1e-4$. The maximum epochs were set to 200 with an early stopping mechanism based on no improvement of validation accuracy over 20 epochs.

The default values of \textbf{BMX} model hyperparameters are summarized as follows.
 
For the multihead attention layer, we selected 51 $\times$ 40 as $V$, and $h = 8$.

The number of CNN layers was three with the number of filters of 8, 16, and 32, respectively. The filter size 
of CNN was 3x3.  The dropout layer was set to 0.5. 
\begin{table*}[htb]
\caption{Summary of BMX Performance against SOTA (\textbf{\%}).}

\label{tab:general performance}
{\small
\centering
\begin{tabular}{|c|c|c|c|a|ccc|}
\hline
\textbf{Subject} &\textbf{DeepSoli}&\textbf{RFWash} &\textbf{Tesla-Rapture}  &\textbf{BMX} & \multicolumn{3}{c|}{\textbf{Improvement Against SOTA}} \\
\cline{6-8}
\textbf{Orientation} & && & \textbf{2 Beams} &DeepSoli &RFWash &Tesla     \\
  \hline
-30$^\circ$&47.2$\pm1.8$&54.8$\pm0.8$&57.1$\pm1.1$ &87.2$\pm4.7$&40.03&32.4&30.1     \\
-15$^\circ$&51.3$\pm0.7$&59.5$\pm7$&59.2$\pm2$&88.2$\pm2.7$&36.9&28.7&29     \\
0$^\circ$&48$\pm0.9$&57.5$\pm5.5$&61.7$\pm6.2$&91.8$\pm1.3$&43.8&34.3&30.2     \\
15$^\circ$&47.2$\pm3.9$&52.9$\pm7.3$&57.2$\pm4.7$&86.9$\pm2.2$&39.7&34&29.3     \\
30$^\circ$&51.7$\pm1.7$&59.6$\pm8.8$&49.8$\pm4.7$&86$\pm3.1$&34.4&26.5&36.3     \\
\hline
\textbf{Average}&49.1&56.8&57 &88&43.1&31.2&31  \\ 
\hline
\end{tabular}
}

\end{table*}

\subsubsection{Metrics and benchmarks}

We used gesture recognition \textit{accuracy} as the performance metric, which is the percentage of test gesture samples that were classified correctly. We first obtained the accuracy for each subject individually using five-fold cross-validation across all positions and environments. We then reported the \textit{average accuracy} over all 10 subjects.

We compared the performance of \textbf{BMX} with three state-of-the-art (SOTA) mmWave-based gesture recognition approaches: DeepSoli~\cite{interactingwithsoli}, RFWash~\cite{rfwash}, and Tesla-Rapture~\cite{tesla}. DeepSoli~\cite{interactingwithsoli} and RFWash~\cite{rfwash} \isn{use RDMs from a single TDM-MIMO channel to recognize gestures with dimensionality of \textbf{Range Bins $\times$ Doppler Bins $\times$ Time Steps }, while Tesla-Rapture}~\cite{tesla} \isn{generates \textit{3D point clouds} from the raw ADC samples obtained from the 12 TDM-MIMO channels and uses them as the input to their deep learning model for classification.}

We implemented these SOTA approaches using the source codes released by their authors\footnote{While no changes were necessary in the Tesla source code, Soli and RFWash codes were adjusted to match the RD dimensions of our dataset.} and evaluated them using the dataset we collected in this work. 
\subsection{Overall Performance}

Table~\ref{tab:general performance} summarizes the gesture recognition accuracy of \textbf{BMX} compared to SOTA approaches, averaged across all subjects and environments. BMX achieved 88\% accuracy on average, a minimum improvement of 31\% (over Tesla) and a maximum improvement of 43\% (over DeepSoli) compared to the SOTA approaches, when two beams are combined and averaged over all possible combinations of two beams with one pointing to the subject.

Additionally, \textbf{BMX} achieved 91.8\% accuracy at the boresight angle and a modest 5\% degradation at the edges of the field of view (-30$^\circ$ and 30$^\circ$). This flexibility allows for excellent user interaction with the system. However, we observed a slight asymmetry in gesture recognition accuracy around the boresight, with positive orientations achieving about 1.5\% less accuracy compared to negative orientations. Our analysis indicated that this might result from occlusions of movements from the dominant hand during some gestures.

\begin{table}[htb]
  \caption{Impact of the number of Beams (\textbf{\%}, one of the beams
  is always pointing to the subject). \textbf{S.O.: Subject Orientation. AVE.: Average}}
  \label{tab:number_beams}
  \centering
\begin{tabular}{|c|c|ccc|}
\hline
\textbf{S. O.} &\textbf{Single} & \multicolumn{3}{c|}{\textbf{Combined Beams}} \\
\cline{3-5}
\textbf{ } &    \textbf{Beam}             &    2   &  3     & 4     \\
  \hline
-30$^\circ$&      72$\pm3.7$         &  87.2 $\pm4.7$     &   86.7$\pm5.8$    &   81.3 $\pm8.2$ \\
 -15$^\circ$&      75.2$\pm5.3$         &  88.2 $\pm2.7$    &   86.6$\pm4.6$    &   83.2 $\pm4.4$  \\
 0$^\circ$&        72$\pm8.87$         &  91.8 $\pm1.3$    &   92.6 $\pm3.2$   &   85.8 $\pm5.9$  \\
 15$^\circ$&       75.3$\pm2.6$         &  86.9 $\pm2.2$    &  88.4 $\pm1$    &   84.2 $\pm9.8$ \\
30$^\circ$&        71.3$\pm1.7$         &  86 $\pm3.1$    &  86.6 $\pm4.6$     &   81.7$\pm8$ \\  
\hline
\textbf{AVE.}&     73.2           &  88     &  88.6     &  83.3  \\  
\hline
\end{tabular}

\end{table}

\subsection{Impact of Number of Beams Combined}
The impact of combining multiple beams on the gesture recognition performance of \textbf{BMX} is shown in Table~\ref{tab:number_beams}. While beam combining is expected to improve performance, it is not immediately clear if further combining would continue to improve performance.

Interestingly, the table shows that the gesture recognition performance saturates with the combination of two beams. There is no further improvement beyond that and in fact, adding more beams can be counterproductive. This indicates that adjacent beams provide wider coverage of the area of interest, leading to better sensing accuracy, while beams pointing far away from the gesture only add to the noise. 

In our analysis of beam selection during the fusion of two beams, we considered the four available beams \{-15, 0, +15, $\pm$30\} and identified three possible combinations, each of which included the beam that aligned with the orientation. We observed that the accuracy of the selection improved by approximately 4\% when we used adjacent beams (e.g., \{-15, 0\} or \{0, +15\}) compared to non-adjacent beams. Using adjacent beams enhances coverage and improves sensing accuracy. Thus, we recommend setting \textbf{BMX} to use two beams: one directed at the subject and the other randomly oriented, without additional overhead.


\subsection{Impact of the Data Augmentation} 
\label{subec:impactDataAugmentataion}

\begin{figure}[t]

    \begin{subfigure}[b]{0.24\textwidth}
        \includegraphics[width=\textwidth]{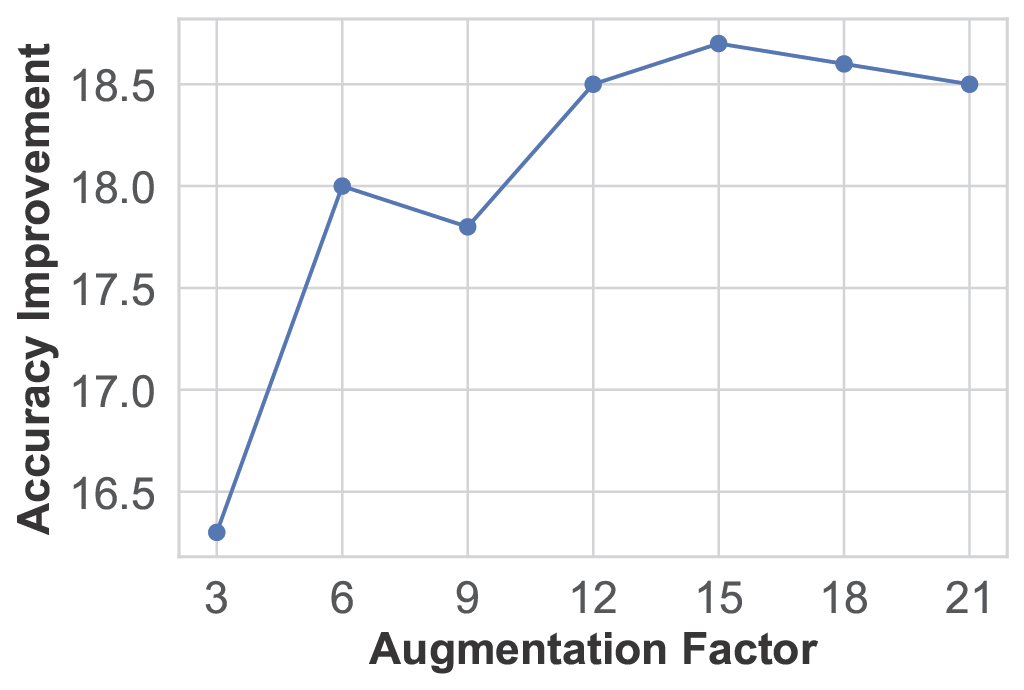}
        \subcaption{}\label{fig:augmentation}
    \end{subfigure}
     \begin{subfigure}[b]{0.22\textwidth}
        \includegraphics[width=\textwidth]{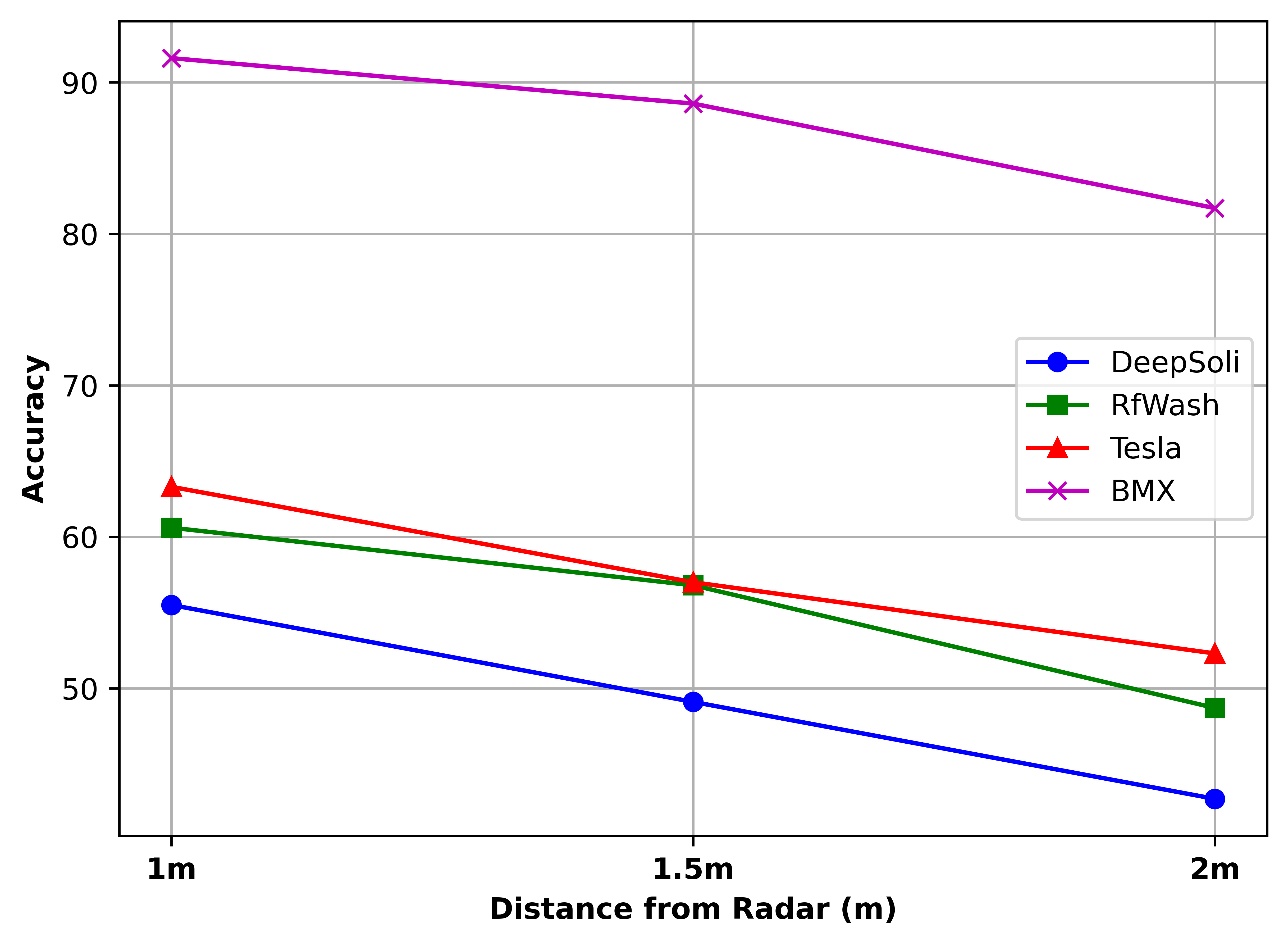}
        \subcaption{}\label{fig:distances}
    \end{subfigure}

    \caption{(a) Impact of Augmentation. (b) Impact of Distance from Radar. Y-axis represents the Avg. Accuracy of beams.}

\end{figure}

To investigate the impact of data augmentation, as introduced in Section~\ref{subsec:dataAugmentation}, we augment the real-world dataset by factors of 3, 6, ..., and 21. Figure~\ref{fig:augmentation} depicts the difference in recognition accuracy between different data augmentation factors and no augmentation.
The results show that our data augmentation algorithm consistently improves accuracy by over 16\% in all cases. The best performance is achieved when the augmentation factor is 15, with an accuracy improvement of over 18.5\%. Therefore, we select an augmentation factor of 15 for the remaining experiments.

\subsection{Impact of Distance}

Our experimentation mainly focused on the 1.5m sensing region, we also tested the performance of BMX along with the benchmarks for distances less or more than our targeted range.
Figure~\ref{fig:distances} summarizes the results when a subject performs gestures
at different distances, i.e., from 1m to 2m.
The results show that, unsurprisingly, the best performance of \textbf{BMX} achieves 
at the closest distance, i.e., 1m, which
is approximately 3\% better than that of 1.5m, which is in turn approximately 7\%
better than that of 2m, due to the 
decrease of SNR. Nevertheless, even at 2m distance, \textbf{BMX} achieves a gesture recognition accuracy rate above 80\%.

Another significant problem is that, the Radar Cross Section (RCS) of a human hand is approximately -35dBsm, while the whole body is -5dBsm~\cite{hugler_geiger_waldschmidt_2016}. Therefore,  a drop in SNR is imminent for micro motion models with longer range.

\begin{figure*}[htb]

 \centering
        \begin{subfigure}[][][t]{0.245\textwidth}
        \includegraphics[width=4cm]{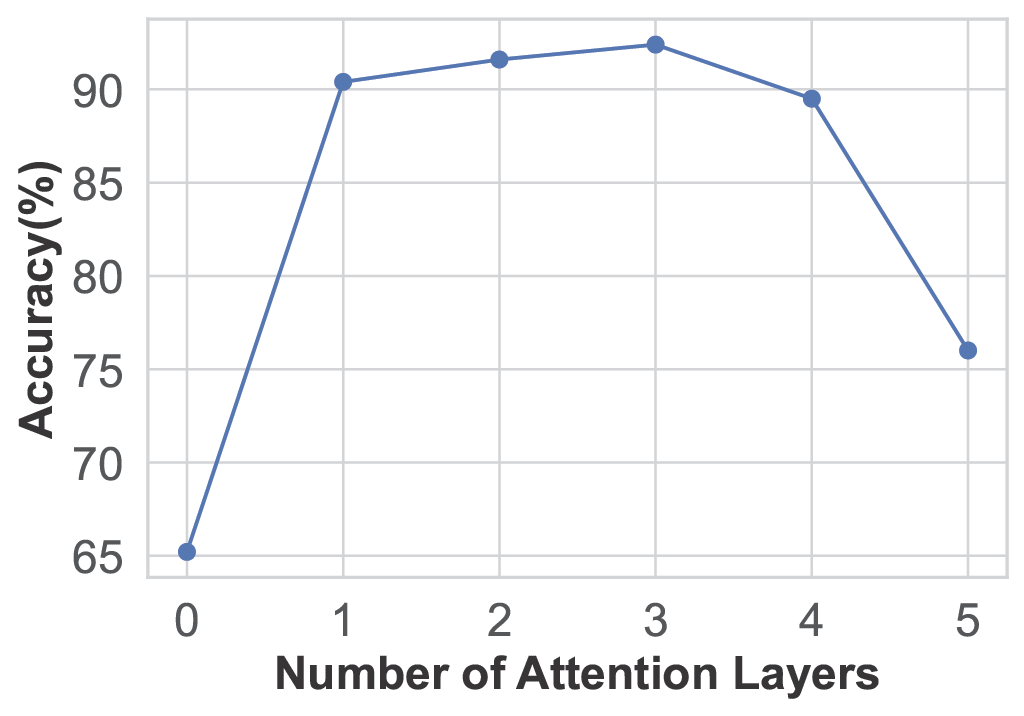}
        \subcaption{Impact of Attention Layers}\label{fig:noattentionLayers}
    \end{subfigure}
    \hfill
    \begin{subfigure}[][][t]{0.245\textwidth}
        \includegraphics[width=4cm]{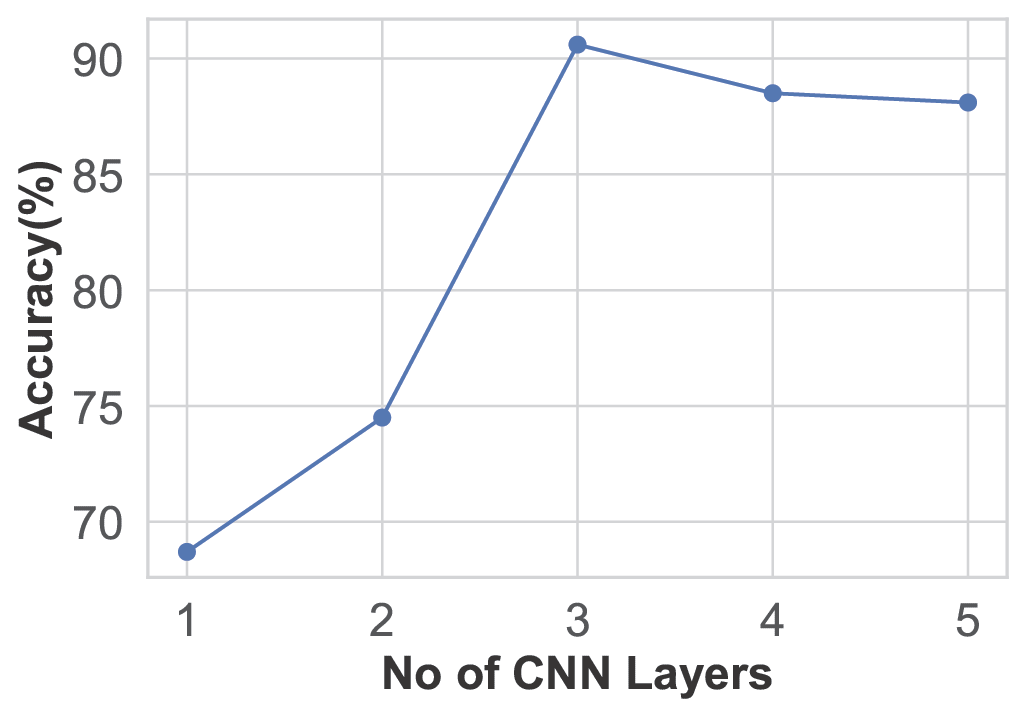}
        \subcaption{Impact of CNN Layers}\label{fig:cnn_layers}
    \end{subfigure}
    \hfill
    \begin{subfigure}[][][t]{0.245\textwidth}
        \includegraphics[width=4cm]{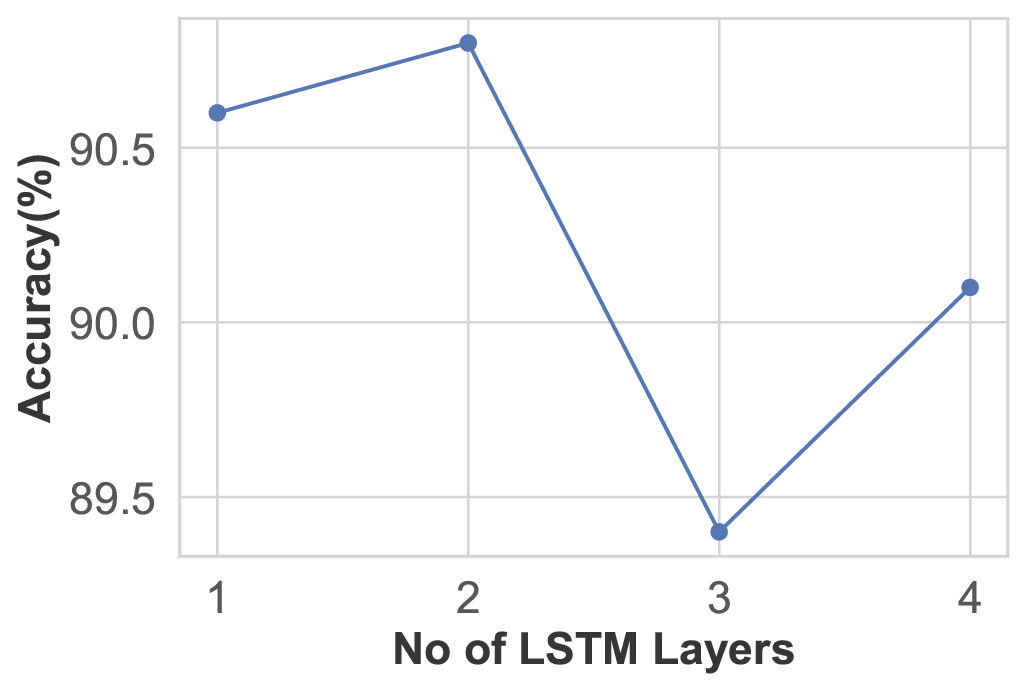}
        \subcaption{Impact of LSTM Layers}\label{fig:lstm_layers}
    \end{subfigure}
    \hfill
    \begin{subfigure}[][][t]{0.245\textwidth}
        \includegraphics[width=4cm]{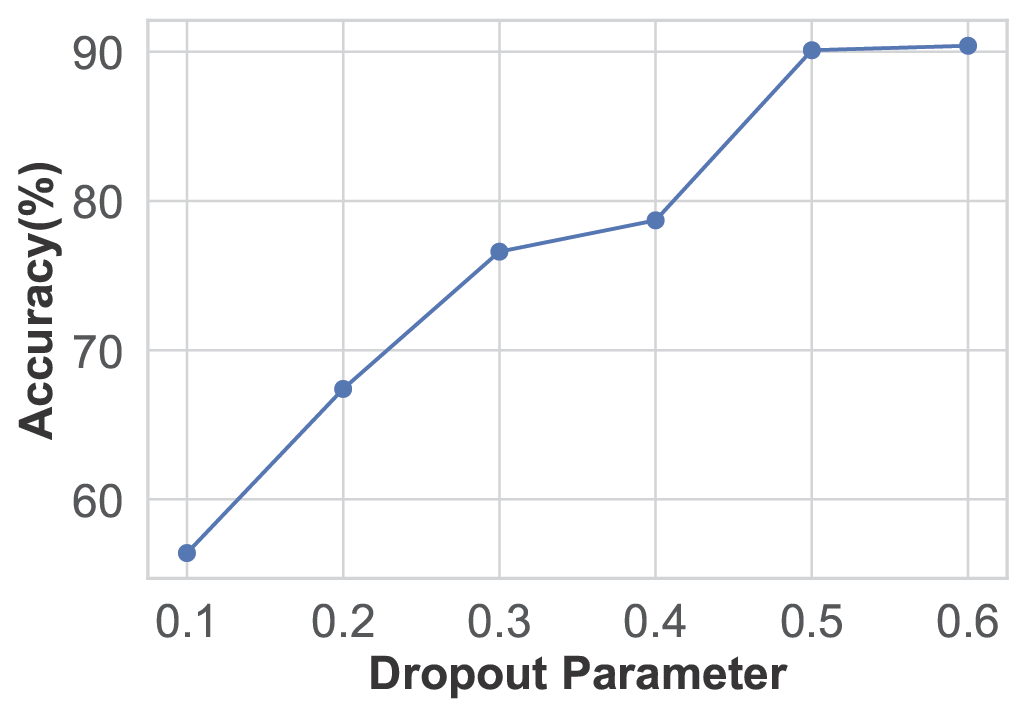}
        \subcaption{Impact of Dropouts}\label{fig:dropouts}
    \end{subfigure}

    \caption{Ablation Study}\label{fig:ablations}

\end{figure*}

\subsection{Impact of Unseen Domains}  

\isn{Evaluating a model's generalization to unseen radio environments, participants, and orientations is crucial for its usability. In the first study, we tested two radio environments and ten subjects, using leave-one-environment-out and leave-one-subject-out methods. The performance of \textbf{BMX} decreased by approximately 6\% for unseen environments and 16\% for unseen subjects, with the latter having a higher impact due to unique gesture performance} (Table~\ref{tab:cross_domain}).

\isn{In the second study, we trained the model using leave-one-orientation-out, achieving 73.5\% accuracy. The primary challenge was the diversity between orientations, especially near the radar's field of view edges.}

\isn{In the final study, we evaluated the model with an unseen gesture, using leave-one-out-gesture training. The average accuracy dropped to 83.6\%, which is a modest degradation given the challenging gesture set}~\cite{pantomime}.

\begin{table}
\centering
\caption{Impact of Unseen Domains}

\small{
\begin{tabular}{|p{2.6cm}|p{1.5cm}|p{1.5cm}|p{1.5cm}|}
\hline

\textbf{Subject's Orientation and Distance} & \textbf{Unseen Environment}& \textbf{Unseen Subject} & \textbf{Unseen Orientation} \\
\hline
\centering
0$^\circ$, 1.5m& 83.0\% & 72.7\% &73.5\%\\

\hline
\end{tabular}
}
\label{tab:cross_domain}

\end{table}

\subsection{Ablation Study}

\textbf{Impact of Number of Attention Layers.} \isn{The attention layers in Figure}~\ref{fig:bmx_deep_model} \isn{are crucial for fusing information from multiple beams but add significant overhead to model parameters. Experiments with varying attention layers} (Figure~\ref{fig:noattentionLayers}) \isn{show that one layer improves performance by 25\%, but additional layers offer marginal gains and decrease performance after four layers. Thus, we use three attention layers.}

\textbf{Impact of the Number of CNN Layers.} \isn{Changing the number of CNN layers from one to five (Figure}~\ref{fig:bmx_deep_model}) \isn{shows considerable improvement up to three layers, with a slight performance drop afterward} (Figure~\ref{fig:cnn_layers}).

\textbf{Impact of the Number of LSTM Layers.} \isn{LSTM layers are vital for learning temporal information from the RDM series} (Figure~\ref{fig:bmx_deep_model}). \isn{Experiments with up to four LSTM layers (Figure} \ref{fig:lstm_layers}) \isn{show that two layers slightly outperform one layer by 0.2\%, but performance fluctuates after two layers. We use one LSTM layer due to added complexity without significant gains.}

\textbf{Impact of the Number of Dropouts.} \isn{A dropout layer at the end of the \textbf{Gesture Sequence Embedding} sub-module} (Figure~\ref{fig:bmx_deep_model}) \isn{addresses overfitting. Experiments with dropout values from 0.1 to 0.5 }(Figure~\ref{fig:dropouts}) \isn{show improved performance up to 0.5, with no further accuracy gains beyond this value. Thus, we use a dropout value of 0.5.}

\textbf{Impact of the Number of Receivers.} \isn{Using a single receiver antenna (Rx1) from the AWR1843 device, we modified BMX to incorporate data from four receiver antennas, resulting in a data structure of} \textbf{Range Bins $\times$ Doppler Bins $\times$ Tx Beams $\times$ Receivers $\times$ Time Steps}. \isn{Combining measurements from four antennas improves performance by approximately 4\% compared to a single receiver, enhancing spatial coverage during gesture execution.}

\subsection{Impact of Maximum Ratio Combining on Performance Benchmarking} \isn{We investigated using Maximum Ratio Combining}~\cite{goldsmith2005wireless} \isn{(MRC) to aggregate data from multiple beams and create a single Range-Doppler Map (RDM) for benchmarking. With a subject orientation of 0°, BMX achieved a performance of 91.8±1.3 using deep learning-based signal combining. However, performance dropped to 86.4±4.5 with traditional signal combining, indicating that data-driven methods used in BMX offer superior gains.

To benchmark state-of-the-art (SOTA) approaches and assess the impact of aggregated data, we used the RDM generated by MRC for BMX, Soli, and RFWash. With a subject orientation of 0°, BMX, Soli, and RFWash achieved 91\%, 48\%, and 57\%, respectively, with performance drops to 86.4\%, 40\%, and 51\%. Tesla-Rapture, benefiting from richer spatial information via TDM-MIMO configuration, cannot be tested with MRC data due to the lack of 3D point clouds in radar beamforming mode. BMX consistently outperforms state-of-the-art (SOTA) methods even with the same input type, primarily due to its specialized deep learning architecture, optimized hyperparameter tuning, and task-specific data augmentation techniques. (.
discussed in the Ablation Study)}
\subsection{Gesture Sequence Order}. \isn{RFWash tracks gesture sequences and compares them to WHO's recommended steps (G1 → G9) but does not strictly enforce order.} \isn{Order enforcement is not the primary focus of BMX. We tested it with random gesture sequences, and BMX performed reliably without any issues.}

\section{Limitations and Future Works}

\isn{\textbf{Sensitivity to the environments and participants}. 
Our primary objective is to enhance low-SNR micro-gesture recognition through beam steering and combination techniques. However, we observed performance variations across different environments, users, and gesture orientations}~\cite{youngjae2020}.\isn{ While leveraging receiver diversity mitigated some of these challenges, it requires extensive training data from a diverse set of users. To address this, our future work will focus on expanding the dataset and refining the model’s adaptability to varying real-world conditions.

\textbf{Realtime Gesture Recognition}. Real-time gesture recognition is crucial for practical deployment, but BMX does not currently support this functionality. Enhancing BMX with real-time capabilities is an important future direction to improve its usability, which we leave for future work.}

\section{Related Work}

\isn{Frequency-Modulated Continuous Wave (FMCW) radar technology has been extensively studied for gesture recognition, leveraging both micro and macro-motion sensing. While several works have explored different techniques, existing approaches suffer from limitations in sensing range, resolution, and adaptability, which BMX addresses by integrating beam steering and deep learning-based signal combination for improved sensing performance.}

\textbf{ Macro Motion Sensing using Radar}.
Soli~\cite{jaime2016,interactingwithsoli} \isn{was among the first radar-based approaches for detecting micro-movements. It has inspired several follow-up works, such as Gait STAR}\cite{8249172}, ThuMouse\cite{9043082}, RadarNet\cite{radarNet}, IndexPen\cite{10.1145/3534601}, Latern\cite{8300506}, and Solids on Soli\cite{solidsonsoli}. \isn{However, these approaches are limited to a sensing range of approximately 30 cm, making them impractical for applications requiring larger coverage.}

\isn{Other works using alternative radar hardware}~\cite{8617375,9778282,9764697,9694576,10.1145/3589645,10.1145/3490099.3511107,liu2022} \isn{have aimed to improve sensing capabilities. Some studies introduce angular sensing enhancements, such as RAI}~\cite{9128573}, \isn{while others propose Pseudo Representational Models (PRM)}~\cite{m-gesture} \isn{to provide spatially richer features that complement Range-Doppler information.}

\textbf{ Micro Motion Sensing using Radar}.
\isn{Recent advancements in 3D point cloud-based gesture recognition have leveraged the Doppler velocity and reflected energy to enhance recognition accuracy}~\cite{pantomime,tesla,Haipeng2021,9043082}. \isn{A key insight from these works is that denser point clouds lead to better recognition due to higher reflected energy, making them beneficial for capturing fine human movements.}

\subsection{Beam Steering and Signal Combining in Radar Sensing}\isn{ Transmit beamforming has primarily been explored in vital sign monitoring to detect subtle physiological movements}~\cite{petropulu2022,10550137,10635792,10683961,10521852,fan_xie_zhou_wang_bu_lu_2024}. \isn{However, these studies focus on static beam directions, meaning the beam remains fixed while detecting signals from a known location. This approach typically requires an initial target localization step to determine the subject’s position before directing the beam and extracting physiological signals. For example,}~\cite{petropulu2022} \isn{determines the target subject’s direction before focusing the beam, making subsequent processing (e.g., vital sign extraction) contingent on precise localization.

In contrast, BMX eliminates the need for explicit localization by sweeping across multiple directions using beam steering diversity. This not only expands the sensing range but also alleviates the target localization burden imposed by static beam approaches. However, this flexibility introduces new challenges, particularly in signal fusion, as multiple beams capture both constructive and destructive multipath reflections. To address this, BMX leverages deep learning-driven beam combination, intelligently fusing information from different angles to enhance gesture recognition accuracy. Unlike prior works that rely on single-beam static transmission, BMX’s multi-angle beamforming enables higher spatial resolution and more robust sensing using commodity hardware.}

\isn{Mechanical beam steering has also been explored to expand the coverage of radar sensing} ~\cite{10878446,argha2024}. \isn{While such approaches effectively extend detection areas, they often require additional hardware modifications or mechanically controlled components, limiting their flexibility for gesture recognition. BMX, in contrast, achieves dynamic beam adaptation purely through software-defined beam control, making it more adaptable to real-world scenarios.}

\isn{Additionally, signal combining has been investigated in wireless sensing to enhance SNR and sensitivity, as demonstrated by DiverSense}~\cite{diverSense} \isn{for Wi-Fi-based sensing. However, previous methods primarily rely on handcrafted signal processing techniques. In contrast, BMX applies deep learning-driven signal combination for mmWave sensing, allowing for automatic pattern learning from constructive and destructive interference, significantly improving gesture recognition robustness.}


\section{Conclusion}

In conclusion, this paper presents a novel mmWave gesture recognition technique, \textbf{BMX}, that improves accuracy in low SNR scenarios and extends its sensing coverage area significantly, which in turn, reduces the system cost to cover a given AoI. By steering mmWave beams towards multiple directions, \textbf{BMX} generates multiple views of gestures that are intelligently combined to enhance gesture classification. Our model features a novel data augmentation algorithm that takes the Doppler shift and multipath into account to improve the generalization of the model. Implementation of \textbf{BMX} on COTS mmWave hardware shows that it outperforms SOTA methods by 31-43\% and achieves 91\% accuracy at boresight by combining only two beams, demonstrating superior gesture classification in low SNR scenarios.

Our results demonstrate the feasibility and effectiveness of mmWave Tx beamforming for gesture recognition, which is an important research direction in the field of mmWave sensing. Future research can explore the applicability of BMX for other gesture sensing applications, such as sign language recognition or human-robot interaction.

\begin{acks}  
We sincerely thank the Australian Research Council (ARC) for their support through the Discovery Project DP220101187. Their generous funding and resources have been instrumental in making this research possible. We also extend our gratitude to all the participants who dedicated their time and effort to the data collection process, contributing significantly to the success of this study.  
\end{acks}

\balance
\bibliographystyle{abbrv}

\bibliography{references}

\end{document}